# First reduced model for integrated computations of helicon wave heating and current drive in magnetic fusion plasmas

Zi-Chen Kan[1], Lei Chang[1,2†], Zhen-Yu Wang[3], Hua-Sheng Xie[4], Ping-Wei Zheng[5], Lai Wei[6], Qi-Bin Luan[7], Xue-Mei Zhai[3], Zhao-Qing Hu[8], Zheng-Xiong Wang[6†], Matthew Hole[9], Zhi-Song Qu[10]

[1]School of Electrical Engineering, Chongqing University, Chongqing 400044, People's Republic of China
[2]Institute of Nuclear Energy and Technology Innovation, Chongqing University, Chongqing 400044, People's Republic of China
[3]Institute of Plasma Physics, HFIPS, Chinese Academy of Sciences, Hefei 230031, People's Republic of China
[4]Hebei Key Laboratory of Compact Fusion, Langfang 065001, People's Republic of China
[5]School of Resources Environment and Safety Engineering, University of South China, Hengyang 421001, People's Republic of China
[6]Key Laboratory of Materials Modification by Laser, Ion, and Electron Beams of the Ministry of Education, School of Physics, Dalian University of Technology, Dalian 116024, People's Republic of China
[7]School of General Education, Dalian University of Technology, Dalian 116024, People's Republic of China
[8]The Southwestern Institute of Physics, Chengdu 610025, People's Republic of China
[9]Mathematical Sciences Institute, Australian National University, ACT, Canberra 0200, Australia
[10]School of Physical and Mathematical Sciences, Nanyang Technological University, Singapore 637371, Singapore

Email: † leichang@cqu.edu.cn (Lei Chang), zxwang@dlut.edu.cn (Zheng-Xiong Wang)

**Abstract:** To enable fast integrated computations of helicon-wave heating and current drive in magnetic fusion plasmas (e.g. tokamaks), this work develops, to our knowledge, the first reduced model that retains only the parallel-electron Landau-damping channel for helicon waves. The model evaluates the wave on the cold-plasma dispersion and adds a single-Landau-pole absorptive correction for Maxwellian electrons, yielding expressions for the damping rate and the local current-drive efficiency. For wave heating, the model is benchmarked against the Chiu-Chan model. The error is decomposed into a truncation contribution from neglected non-Landau damping channels and an approximation contribution within the retained Landau channel. A scan of approximately $1.6\times10^6$ samples from EAST, HL-3 (formerly HL-2M), DIII-D and KSTAR shows that the Landau parameter and the electron beta jointly control the validity boundary. Within the restricted validity window, the errors from all four devices collapse onto a single curve that can be fitted by a fifth-order polynomial. The resulting empirical correction substantially extends the model applicability and is verified with ITER-like and BEST-like extrapolation cases. Above the lower-hybrid frequency, both the uncorrected and corrected models show abrupt error growth correlated with topological changes in the cold-dispersion root structure. This behaviour indicates that the failure is associated with the validity boundary of the cold-dispersion WKB framework. The corrected heating model is then coupled to a reduced local current-drive source and benchmarked against the Ehst-Karney model. Within the heating-model validity domain and under a Landau-channel-consistent comparison, the median deviation of the reduced current-drive efficiency is approximately 10.8%. A post-processing test with published CFETR absorbed-power profiles further shows that the model can reasonably reconstruct the radial shape of helicon-driven current-density profiles. A first-order single-pass absorption analysis shows that the empirical correction reduces error transfer in the weak-absorption limit, while absorption saturation suppresses error transfer in the strong-absorption limit.

## NOMENCLATURE

| Symbol | Description |
|---|---|
| $B_0$ | Equilibrium magnetic field |
| $n_e$ | Electron density |
| $T_e$ | Electron temperature |
| $a$ | Minor radius |
| $R_0$ | Major radius |
| $\epsilon$ | Inverse aspect-ratio parameter |
| $k_\parallel$ | Parallel wavenumber |
| $k_\perp$ | Perpendicular wavenumber |
| $k_{\perp R}$ | Real part of the perpendicular wavenumber in the reference model |
| $k_{\perp i}$ | Imaginary part of the perpendicular wavenumber; local spatial damping rate |
| $k_{\perp i}^{\mathrm{red}}$ | Reduced-model damping rate |
| $k_{\perp i}^{\mathrm{ref}}$ | Reference damping rate |
| $k_{\perp i}^{\mathrm{corr}}$ | Corrected reduced-model damping rate |
| $N_\parallel$ | Parallel refractive index |
| $N_\perp$ | Perpendicular refractive index |
| $\xi_e$ | Electron Landau parameter |
| $\beta_e$ | Electron beta |
| $Z(\xi_e)$ | Fried-Conte plasma dispersion function |
| $Z_R(\xi_e)$ | Hermitian part of $Z(\xi_e)$ |
| $Z_I(\xi_e)$ | Anti-Hermitian Landau part of $Z(\xi_e)$ |
| $\eta_L$ | Landau-dominance ratio |
| $\rho$ | Closure-quality factor for the retained Landau channel |
| $\rho_{\mathrm{fit}}(Z_R)$ | Empirical correction factor |
| $\delta_k^a$ | Relative damping-rate error of the reduced model |
| $\delta_k^{\mathrm{corr}}$ | Relative damping-rate error after correction |
| $P_{\mathrm{abs}}$ | Absorbed wave power density or prescribed absorbed-power source |
| $J_{\mathrm{CD}}$ | Driven current-density source |
| $\eta_{\mathrm{CD}}$ | Local current-drive efficiency per unit absorbed power |
| $\eta_{\mathrm{CD}}^{\mathrm{red}}$ | Reduced local current-drive efficiency |
| $\eta_{\mathrm{FB}}$ | Fisch-Boozer current-drive efficiency factor |
| $w_{\mathrm{EK}}$ | Ehst-Karney velocity parameter |
| $\mathcal{R}_{\mathrm{rel}}$ | Weakly relativistic correction factor |
| $T_{\mathrm{trap}}$ | Trapped-particle correction factor |
| $f_t^{\mathrm{eff}}$ | Effective trapped-particle fraction |
| $\tau$ | Single-pass optical depth |
| $L$ | Effective single-pass propagation length |
| $f_{\mathrm{abs}}^{\mathrm{ref}}$ | Single-pass absorbed fraction |
| $\delta_{\mathrm{CD}}$ | Relative current-drive source error propagated from the damping-rate error |
| $\delta_{\eta_{\mathrm{A}}}$ | Relative error of the local current-drive efficiency against the LD-channel reference |
| $\mathcal{A}$ | First-order absorption-error transfer factor |

## 1 Introduction

Helicon waves, traditionally treated as a high-harmonic branch of the fast wave in tokamak research, have recently attracted renewed interest as a candidate for radio-frequency (RF) wave heating and current drive in high-density tokamak plasmas[1-4]. Under the high-density, high-field conditions expected in next-generation tokamaks, RF wave heating and current-drive schemes offer favourable wave accessibility and flexible control over the deposition location[5-7]. Compared with other RF waves, helicon waves propagate and penetrate more effectively in high-density plasmas, which makes them a promising candidate for mid-radius off-axis current drive[1,2,8]. However, helicon-wave propagation, single-pass absorption, and current-drive efficiency are highly sensitive to the launched frequency, the antenna parallel-refractive-index ( $N_{\parallel}$ ) spectrum, and the background plasma profiles[9]. Both physics assessment and engineering design therefore require large-scale parameter scans and integrated modelling, which in turn demand a model that combines adequate physical fidelity with low computational cost[10-12].

These two requirements compete in practice, and existing approaches occupy different aspects and regimes on the resulting trade-off. Full-wave codes such as AORSA resolve propagation, mode conversion, diffraction, and single-pass absorption, and are well suited to high-fidelity validation, but their cost makes large scans over frequency, antenna spectrum, and profile impractical[2,13-18]. Ray-tracing codes such as GENRAY and BORAY instead follow characteristic rays governed by a local dispersion relation and represent damping through the imaginary part of the complex wavenumber[19-24]. They strike a useful balance between fidelity and cost and are well suited to rapid parametric studies in engineering-oriented design [19–21]. The local damping model that supplies this imaginary wavenumber at each ray point therefore controls both the accuracy and the cost of such scans. Existing models, however, were developed for broader fast-wave applications rather than for a helicon-wave regime governed by a single dominant damping channel. The error introduced by such a single-channel reduction, and the parameter range over which it remains reliable, have not been quantified.

For helicon waves, this damping model is usually built from the cold fast-wave branch with hot-plasma corrections that retain several damping mechanisms, principally electron Landau damping (ELD) and transit-time magnetic pumping (TTMP), with the Chiu-Chan model as a representative example[1,25-27]. Reduced kinetic descriptions of this kind have a long history in plasma modelling, from fluid-moment closures for turbulence and collisionless MHD[28-31] to the tractable damping-rate and current-drive formulae of early fast-wave theory[25,32]. These formulations were developed primarily for the fast-wave regime and therefore retain multiple damping mechanisms. In the helicon operating window considered here, ELD is the dominant absorption mechanism. This dominance suggests that the damping model can be reduced to a single retained channel. However, no reduced model tailored to this ELD-dominated helicon window has yet been systematically formulated, benchmarked, and given a quantitative validity range.

The present work addresses this gap. We develop a reduced model for helicon-wave heating and current drive. The model retains parallel ELD as the only damping channel and evaluates the wave response on the cold-dispersion root. It therefore avoids solving a thermally corrected dispersion relation while yielding analytical expressions for both the damping rate and the local current-drive efficiency. The aim is to retain computational speed while keeping the error structure interpretable and the validity domain explicit. For heating, the Chiu-Chan model is used as the benchmark[25]. The model error is decomposed into two parts: the error caused by neglecting non-Landau damping channels, and the approximation error within the retained Landau damping channel. The cross-device validity domain of the reduced model is then identified through parameter scans based on four representative device parameter sets, namely EAST[33] and HL-3 (formerly HL-2M)[34] in China, DIII-D[35] in the United States, and KSTAR[36] in the Republic of Korea. Within this domain, a self-contained empirical correction is constructed and tested on ITER-like and BEST-like hold-out cases. The analysis also identifies a failure boundary associated with topological changes in the cold-plasma dispersion roots. A first-order single-pass-absorption analysis then quantifies how damping-rate errors propagate into the absorbed power and the current-drive source. For current drive, the corrected heating model is coupled to a reduced local source term. The resulting local current-drive efficiency is benchmarked against the Ehst-Karney model within the heating-validity domain. A post-processing test against a published CFETR case further checks whether the reduced source can reproduce the radial shape of the current-density profile. The study focuses on wave damping, power deposition, and current drive in the closed-flux-surface region. Scrape-Off Layer (SOL) physics and antenna-coupling effects lie outside its scope[9]. The remainder of the paper is organized as follows. Section 2 develops the reduced model. Section 3 describes the benchmark setup. The heating and current-drive results are presented in Sections 4 and 5. The main conclusions are summarized in Section 6.

# 2 Theoretical Model

### 2.1 Reduced model of helicon wave heating

In a local field-aligned frame, the cold Stix dielectric tensor is written as[37]:

$$\boldsymbol{\varepsilon}_c = \begin{pmatrix} S_c & -iD_c & 0 \\ iD_c & S_c & 0 \\ 0 & 0 & P_c \end{pmatrix} \tag{1}$$

with

$$S_c = 1 - \sum_s \frac{\omega_{ps}^2}{\omega^2 - \omega_{cs}^2} \qquad D_c = \sum_s \frac{\omega_{cs}}{\omega} \frac{\omega_{ps}^2}{\omega^2 - \omega_{cs}^2} \qquad P_c = 1 - \sum_s \frac{\omega_{ps}^2}{\omega^2} \tag{2}$$

Using the parallel and perpendicular refractive indices, $N_\parallel = k_\parallel c/\omega$ and $N_\perp = k_\perp c/\omega$, the cold-plasma dispersion relation obtained from Maxwell's equations becomes:

$$F(\omega, k_\parallel^2, k_\perp^2) = S_c \frac{k_\perp^4 c^4}{\omega^4} - \left[ (S_c + P_c)\left( S_c - \frac{k_\parallel^2 c^2}{\omega^2} \right) - D_c^2 \right] \frac{k_\perp^2 c^2}{\omega^2} + P_c \left[ \left( S_c - \frac{k_\parallel^2 c^2}{\omega^2} \right)^2 - D_c^2 \right] \tag{3}$$

In this representation, the $P$ element governs the dielectric response along the equilibrium magnetic field. Since helicon waves are launched almost parallel to $\mathbf{B}_0$ and their dominant dissipation channel is ELD, the thermal correction is retained only in the parallel component. The modified dielectric tensor then becomes:

$$\boldsymbol{\varepsilon}_{\text{eff}}^{(1)} = \begin{pmatrix} S_c & -iD_c & 0 \\ iD_c & S_c & 0 \\ 0 & 0 & P_c + \delta P_h^{(H)} + i\delta P_h^{(A)} \end{pmatrix} \tag{4}$$

Here, $\delta P_h^{(H)}$ and $\delta P_h^{(A)}$ are the Hermitian and anti-Hermitian parts of the parallel thermal correction, respectively. The Hermitian contribution mainly shifts the real propagation root and is neglected in the reduced model. Retaining only the absorptive contribution gives:

$$\begin{aligned} \boldsymbol{\varepsilon}_{\text{eff}}^{(1)} &= \boldsymbol{\varepsilon}_c + i\boldsymbol{\varepsilon}_{aL} \\ &= \begin{pmatrix} S_c & -iD_c & 0 \\ iD_c & S_c & 0 \\ 0 & 0 & P_c + i\delta P_h^{(A)} \end{pmatrix} \end{aligned} \tag{5}$$

The sign of the imaginary part is fixed by the Fourier time convention adopted throughout this work, chosen such that local absorption is positive. It is convenient to introduce:

$$X \equiv N_\perp^2 \qquad \Gamma \equiv S_c - N_\parallel^2 \qquad \Delta \equiv \Gamma^2 - D_c^2 \tag{6}$$

The thermally corrected dispersion function can then be written as:

$$\Omega_{\text{eff}}^{(1)}(X) = S_c X^2 + \left[ -(S_c + P_c + i\delta P_h^{(A)})\Gamma + D_c^2 \right] X + (P_c + i\delta P_h^{(A)})\Delta \tag{7}$$

This form separates the dispersion function into a cold real part and a small absorptive correction:

$$\Omega_{\text{eff}}^{(1)}(X) = \Omega_c(X) + i\delta P_h^{(A)} M(X) \tag{8}$$

with

$$M(X) \equiv \Delta - \Gamma X = S_c\left(N_\parallel^2 - N_\perp^2\right) + N_\parallel^2 N_\perp^2 - D_c^2 \tag{9}$$

Under the weak-damping approximation, we write the propagating root as $X = X_c + \mathrm{i}X_i$ , where $X_c$ solves $\Omega_c(X_c) = 0$ and $|X_i| \ll |X_c|$. Expanding $\Omega_c$ to first order around $X_c$ gives:

$$\Omega_c(X) \approx \Omega_c(X_c) + \left(\frac{\partial \Omega_c}{\partial X}\right)_{X_c} (iX_i) \tag{10}$$

After substituting $\Omega_c(X_c) = 0$, Eq.(10) reduces to

$$\Omega_{\text{eff}}^{(1)}(X) \approx iX_i \left(\frac{\partial \Omega_c}{\partial X}\right)_{X_c} + i\delta P_h^{(A)} M(X_c) \tag{11}$$

By setting $\Omega_{\text{eff}}^{(1)}(X) = 0$, we obtain the imaginary part of the root:

$$X_i = -\frac{\delta P_h^{(A)} M(X_c)}{\left(\frac{\partial \Omega_c}{\partial X}\right)_{X_c}} \tag{12}$$

When writing $n_\perp = n_{\perp c} + in_{\perp i}$, the weak-damping limit gives $X_i \simeq 2n_{\perp c} n_{\perp i}$, so that

$$N_{\perp i} = \frac{X_i}{2N_{\perp c}} = -\frac{\delta P_h^{(A)} M(X_c)}{2N_{\perp c}\left(\frac{\partial \Omega_c}{\partial X}\right)_{X_c}} \tag{13}$$

Converting to the perpendicular wavenumber $k_\perp = (\omega / c) n_\perp$, we obtain:

$$k_{\perp i}^{\text{red}} = \frac{\omega}{c} N_{\perp i} = \frac{\omega}{c} \frac{\delta P_h^{(A)} M(X_c)}{2N_{\perp c}\left(\frac{\partial \Omega_c}{\partial X}\right)_{X_c}} \tag{14}$$

The parallel thermal correction is defined as:

$$\delta P_h \equiv \varepsilon_\parallel^{\text{hot}} - P_c \tag{15}$$

Retaining only the $n = 0$ electron parallel resonance, the parallel electron susceptibility takes the standard form:

$$\begin{aligned} \chi_{e\parallel} &= \frac{2\omega_{pe}^2}{k_\parallel^2 v_{te}^2}\left[1 + \xi_e Z(\xi_e)\right] \\ \xi_e &= \frac{\omega}{k_\parallel v_{te}} \end{aligned} \tag{16}$$

where $Z(\xi)$ is the Fried-Conte plasma dispersion function[38]. The parallel hot dielectric response therefore reads:

$$\varepsilon_\parallel^{\text{hot}} = 1 + \chi_{e\parallel} = 1 + \frac{2\omega_{pe}^2}{k_\parallel^2 v_{te}^2}\left[1 + \xi_e Z(\xi_e)\right] \tag{17}$$

The combination of Eqs.(2), (15) and (17) gives:

$$\begin{aligned} \delta P_h &= \delta P_h^{(H)} + i\delta P_h^{(A)} \\ &= \frac{2\omega_{pe}^2}{k_\parallel^2 v_{te}^2}\left[1 + \xi_e Z(\xi_e)\right] + \frac{\omega_{pe}^2}{\omega^2} \end{aligned} \tag{18}$$

By using the Landau prescription, we get the imaginary part of $Z(\xi_e)$ for real $\xi_e$ [37]:

$$\begin{aligned} &Z(\xi_e) = Z_R(\xi_e) + iZ_I(\xi_e) \\ &\Re\left(Z_I(\xi_e)\right) = \sqrt{\pi} e^{-\xi_e^2} \end{aligned} \tag{19}$$

As a result, the absorptive part of Eq.(18) becomes:

$$\delta P_h^{(A)} \equiv 2\sqrt{\pi} \frac{\omega_{pe}^2}{k_\parallel^2 v_{te}^2} \xi_e e^{-\xi_e^2} \tag{20}$$

which is taken as the magnitude of the Landau-type parallel thermal correction.

Substituting Eqs.(9), (14) and (20) into the weak-damping expression gives the reduced spatial damping rate:

$$k_{\perp i}^{\text{red}} = -\sqrt{\pi} \frac{\omega^3 \omega_{pe}^2}{c^2 k_{\perp c} k_\parallel^3 v_{te}^3} \exp\left(-\frac{\omega^2}{k_\parallel^2 v_{te}^2}\right) \frac{\left(S_c - \frac{c^2 k_\parallel^2}{\omega^2}\right)^2 - D_c^2 - \left(S_c - \frac{c^2 k_\parallel^2}{\omega^2}\right)\frac{c^2 k_{\perp c}^2}{\omega^2}}{2S_c \frac{c^2 k_{\perp c}^2}{\omega^2} - (S_c + P_c)\left(S_c - \frac{c^2 k_\parallel^2}{\omega^2}\right) + D_c^2} \tag{21}$$

Eq.(21) expresses the spatial damping rate of the helicon wave branch entirely in terms of cold-plasma quantities evaluated on the real-root surface $X = X_c$, together with a single Maxwellian Landau factor. This provides a compact and numerically efficient surrogate for the full hot dispersion analysis.

**2.2 Reduced model of helicon current drive**

The reduced damping model derived above provides the local spatial damping rate of the helicon branch. To extend the model from heating to current-drive estimates, the absorbed wave power is coupled to a reduced local current-drive efficiency. The resulting source term plays the same local power-to-current conversion role as the Ehst-Karney parameterization used in ray-tracing computations[32].

When the empirically corrected reduced damping rate is used, the reduced current-drive source is written as:

$$J_{\mathrm{CD}}^{\mathrm{red}} = \eta_{\mathrm{CD}}^{\mathrm{red}} P_{\mathrm{abs}}^{\mathrm{red}} \tag{22}$$

where $\eta_{\mathrm{CD}}^{\mathrm{red}}$ is the local current-drive efficiency per unit absorbed power. It is decomposed as:

$$\eta_{\mathrm{CD}}^{\mathrm{red}} = \eta_{\mathrm{FB}}^{\mathrm{red}} \mathcal{R}_{\mathrm{rel}} \mathcal{T}_{\mathrm{trap}} \tag{23}$$

Consistent with the heating model, no explicit TTMP current-drive channel is retained. The current-drive efficiency is therefore determined by the parallel phase velocity of the Landau-resonant electrons. For a given $N_{\parallel}$, the resonant velocity is:

$$u_{\parallel} = \frac{\omega}{k_{\parallel}} = \frac{c}{N_{\parallel}} \tag{24}$$

After absorbing wave energy, these resonant electrons generate a net parallel current during collisional relaxation. Following the Fisch-Boozer high-phase-velocity limit, the lowest-order local current-drive efficiency is[39]:

$$\eta_{\mathrm{FB}}^{\mathrm{LD}} = \frac{4w_{\mathrm{EK}}^2}{5+Z_{\mathrm{eff}}} \tag{25}$$

with

$$w_{\mathrm{EK}} = \frac{\sqrt{2}c}{N_{\parallel} v_{te}} \tag{26}$$

This is the leading high-phase-velocity contribution. Under practical helicon-wave parameters, however, $w_{\mathrm{EK}}$ is not always sufficiently large, and the bare Fisch–Boozer term may underestimate the current-drive efficiency. The finite-$w_{\mathrm{EK}}$ correction from the LD branch of the Ehst-Karney model is therefore included[32]:

$$\eta_{\mathrm{FB}}^{\mathrm{red}} = \frac{K_{\mathrm{LD}}}{w_{\mathrm{EK}}} + D_{\mathrm{LD}} + \frac{4w_{\mathrm{EK}}^2}{5+Z_{\mathrm{eff}}} \tag{27}$$

with

$$K_{\mathrm{LD}} = \frac{3}{Z_{\mathrm{eff}}}, \qquad D_{\mathrm{LD}} = \frac{3.83}{Z_{\mathrm{eff}}^{0.707}} \tag{28}$$

For helicon-wave parameters, $n_{\parallel}$ is typically of order a few, so that $u_{\parallel}/c$ is not completely negligible. A weakly relativistic correction is included as[40]:

$$\mathcal{R}_{\mathrm{rel}} = \frac{1}{\gamma_w^2 \left[1 + w^2 H(Z_{\mathrm{eff}})\right]} \tag{29}$$

with

$$w = \frac{u_{\parallel}}{c} = \frac{1}{N_{\parallel}} \tag{30}$$

$$\gamma_w = \frac{1}{\sqrt{1-w^2}} \tag{31}$$

$$H(Z_{\mathrm{eff}}) = \frac{3}{2}\frac{3+Z_{\mathrm{eff}}}{5+Z_{\mathrm{eff}}} \tag{32}$$

For $w \ll 1$, this correction reduces to:

$$\mathcal{R}_{\mathrm{rel}} \simeq 1 - w^2\left[1 + H(Z_{\mathrm{eff}})\right] + O(w^4) \tag{33}$$

The correction is smaller than unity and represents the reduction of the current-drive efficiency due to weakly relativistic effects. Magnetic trapping is included through a trapped-particle correction[41,42]:

$$\mathcal{T}_{\text{trap}} = 1 - \frac{f_t^{\text{eff}}}{1+\sqrt{(1+Z_{\text{eff}})/2}} \tag{34}$$

In the absence of a full equilibrium-dependent trapped-particle computation, the effective trapped-particle fraction is approximated by the large-aspect-ratio expression[41,42]:

$$f_t^{\text{eff}} = 1 - \frac{(1-\epsilon)^2}{\left(1+1.46\sqrt{\epsilon}\right)\sqrt{1-\epsilon^2}} \qquad \epsilon = \frac{a}{R_0} \tag{35}$$

This factor reduces the current generated per unit absorbed power, because trapped electrons do not contribute efficiently to the net parallel current. It modifies the current-drive efficiency but not the wave absorption itself. Combining the Fisch-Boozer efficiency, the weakly relativistic correction, and the trapped-particle correction gives the reduced local current-drive efficiency:

$$\eta_{\text{CD}}^{\text{red}} = \left(\frac{3.0}{Z_{\text{eff}}\, w_{\text{EK}}} + \frac{3.83}{Z_{\text{eff}}^{0.707}} + \frac{4w_{\text{EK}}^2}{5+Z_{\text{eff}}}\right)\cdot\left[\frac{1-\frac{1}{N_\parallel^2}}{1+\frac{1}{N_\parallel^2}\left(\frac{3}{2}\cdot\frac{3+Z_{\text{eff}}}{5+Z_{\text{eff}}}\right)}\right]\cdot\left(\frac{\sqrt{(1+Z_{\text{eff}})/2}}{1+\sqrt{(1+Z_{\text{eff}})/2}} + \frac{(1-a/R_0)^{3/2}}{\left(1+1.46\sqrt{a/R_0}\right)\left(1+\sqrt{(1+Z_{\text{eff}})/2}\right)}\right) \tag{36}$$

Finally, multiplying this local efficiency by the reduced absorbed power gives the reduced helicon current-drive source:

$$J_{\text{CD}}^{\text{red}} = \left(\frac{3.0}{Z_{\text{eff}}\, w_{\text{EK}}} + \frac{3.83}{Z_{\text{eff}}^{0.707}} + \frac{4w_{\text{EK}}^2}{5+Z_{\text{eff}}}\right)\cdot\left[\frac{1-\frac{1}{N_\parallel^2}}{1+\frac{1}{N_\parallel^2}\left(\frac{3}{2}\cdot\frac{3+Z_{\text{eff}}}{5+Z_{\text{eff}}}\right)}\right]\cdot\left(\frac{\sqrt{(1+Z_{\text{eff}})/2}}{1+\sqrt{(1+Z_{\text{eff}})/2}} + \frac{(1-a/R_0)^{3/2}}{\left(1+1.46\sqrt{a/R_0}\right)\left(1+\sqrt{(1+Z_{\text{eff}})/2}\right)}\right)\cdot P_{abs} \tag{37}$$

Eq.(37) gives the direct expression for evaluating the reduced current-drive source once the model is embedded in a ray-tracing calculation. Since no ray-tracing simulation is introduced in the present study, however, the main benchmark of the current-drive model is restricted to the local efficiency in Eq.(36).

### 2.3 Reference model and error analyses

#### 2.3.1 Regarding wave heating model

Existing ray-tracing simulations of helicon-wave heating and current drive commonly use the Chiu-Chan model to evaluate the local power deposition. Within the helicon wave operating window, TTMP and ELD constitute the dominant kinetic damping channels. In this regime, the all-hot dielectric tensor model reduces numerically to the Chiu-Chan model, which makes the latter a natural physical reference. Appendix A provides a consistency check against the all-hot tensor complex-root calculation. In the Chiu-Chan model, the perpendicular damping rate is written as [25]:

$$k_{\perp i}^{\text{ref}} = \frac{\sqrt{\pi}}{4}\beta_e \xi_e e^{-\xi_e^2} G k_{\perp R} \tag{38}$$

$$\beta_e = \frac{2\mu_0 n_e T_e}{B^2} = \frac{\omega_{pe}^2 v_{te}^2}{\omega_{ce}^2 c^2} \tag{39}$$

Here, G is the wave damping factor and $k_{\perp R}$ denotes the real part of the perpendicular wavenumber obtained from the dispersion relation that retains the Hermitian electron parallel thermal correction. The total damping factor admits a natural decomposition into a non-Landau contribution $G_1$ and a parallel ELD contribution $G_2$:

$$\begin{aligned} G &= G_1 + G_2 \\ G_1 &= \frac{n_\parallel^2 - S}{A_2} \\ G_2 &= \left[1 + \frac{\omega_{pe}^2 \operatorname{Re}(Y)}{\omega_{ce}^2 |Y|^2}\frac{1}{n_\parallel^2 - S}\right]\left(\frac{m_e c^2}{kT_e}\right)\frac{\omega_{ce}^2 D^2}{\omega^2 |P_{\text{eff}}|^2}\frac{1}{A_1 A_2} \end{aligned} \tag{40}$$

with

$$A_1 = n_\parallel^2 - S + \frac{\omega_{pe}^2 \mathrm{Re}(Y)}{\omega_{ce}^2 |Y|^2} \frac{n_\parallel^2}{n_\parallel^2 - S}$$
$$A_2 = n_\parallel^2 - S + \frac{\omega_{pe}^2 \mathrm{Re}(Y)}{\omega_{ce}^2 |Y|^2} \frac{n_\parallel^2 + S}{n_\parallel^2 - S} \tag{41}$$

$$Y = 2\xi_e^2 \left[1 + \xi_e Z(\xi_e)\right]$$
$$Z(\xi_e) = i\pi^{1/2} \exp\left(-\xi_e^2\right) - 2\xi_e N(\xi_e) \tag{42}$$
$$N(\xi_e) = \left[\exp\left(-\xi_e^2\right) / \xi_e\right] \int_0^{\xi_e} \exp\left(t^2\right) dt$$

$$\xi_e = v_\parallel / v_{te} = \omega / \left(k_\parallel v_{te}\right) = c / \left(N_\parallel v_{te}\right) \tag{43}$$

For a quantitative comparison between the reduced model and the Chiu-Chan reference model within the helicon wave operating window, we make the $G_1 + G_2$ split explicit:

$$k_{\perp i}^{\mathrm{ref}} = \frac{\sqrt{\pi}}{4} \beta_e \xi_e e^{-\xi_e^2} k_{\perp R} (G_1 + G_2) \tag{44}$$

and define the Landau-dominance ratio:

$$\eta_L = \frac{G_2}{G_1 + G_2} \tag{45}$$

Turning to the reduced model derived, substitution of Eq.(9), (14), and (20) yields the result:

$$k_{\perp i}^{\mathrm{red}} = \sqrt{\pi} \frac{\omega_{pe}^2}{\omega^2} \xi_e^3 e^{-\xi_e^2} k_{\perp C} \cdot \frac{|M(X_c)|}{\left|\frac{\partial \Omega_c}{\partial X}\right|_{X_C}} \tag{46}$$

For a like-for-like comparison with the Chiu-Chan expression, we recast Eq.(46) into the same prefactor structure as the reference model:

$$k_{\perp i}^{\mathrm{red}} = \frac{\sqrt{\pi}}{4} \beta_e \xi_e e^{-\xi_e^2} k_{\perp R} \tilde{G} \tag{47}$$

with

$$\begin{aligned} \tilde{G} &= \frac{4\omega_{pe}^2 \xi_e^2}{\beta_e \omega^2} \cdot \frac{|M(X_c)|}{\left|\frac{\partial \Omega_c}{\partial X}\right|_{X_C}} \cdot \frac{k_{\perp C}}{k_{\perp R}} \\ &= \frac{4\omega_{ce}^2 \xi_e^2 c^2}{v_{te}^2 \omega^2} \cdot \frac{|M(X_c)|}{\left|\frac{\partial \Omega_c}{\partial X}\right|_{X_C}} \cdot \frac{k_{\perp C}}{k_{\perp R}} \end{aligned} \tag{48}$$

The error between the two models is defined as:

$$\delta_k^a = \frac{k_{\perp i}^{\mathrm{red}}}{k_{\perp i}^{\mathrm{ref}}} - 1 = \frac{\tilde{G}}{G_1 + G_2} - 1 \tag{49}$$

Substituting the Landau-dominance ratio in Eq.(45) into this error definition gives:

$$\begin{aligned} \delta_k^a &= \eta_L \cdot \frac{\tilde{G}}{G_2} - 1 \\ &= -(1 - \eta_L) + \eta_L (\rho - 1) \end{aligned} \tag{50}$$

$$\rho = \frac{\tilde{G}}{G_2} \tag{51}$$

The model error therefore decomposes into two contributions: a channel truncation error $-(1-\eta_L)$ caused by neglecting non-Landau damping, and an intra-channel approximation error $\eta_L(\rho - 1)$ arising from the reduced model's treatment of the Landau channel itself. The reduced model is consequently expected to be quantitatively accurate only when both $\rho$ and $\eta_L$ approach unity.

A key structural difference between the reduced model and the Chiu-Chan model is that they evaluate the damping rate at different real dispersion roots. In the Chiu-Chan model, $k_{\perp R}$ is obtained by solving a dispersion relation that includes the Hermitian thermal correction. The parallel dielectric component takes the form $P_c + \delta P_h^{(H)}$, and it depends on the real part of the Fried-Conte function. By contrast, the reduced model evaluates the Landau damping correction at the real root $X_c$ of the cold-dispersion root from Equation $\Omega_c(X) = 0$. This difference translates into a measurable difference in local damping-evaluation cost in ray-tracing applications, as shown in appendix A. For the reduced model, the required cold real wavenumber is a local quantity already computed during ray propagation. In contrast, $k_{\perp R}$ in the Chiu-Chan model cannot be directly replaced by the cold real root supplied by the ray tracing.

In the reduced model, the Landau term retains only the imaginary part of the plasma dispersion function, $Z_{\rm I}(\xi_e) = \sqrt{\pi}\exp(-\xi_e^2)$. This factor has an elementary form. It is evaluated at the cold real root, $k_{\perp 0}$, which is already provided by the ray-tracing calculation. Therefore, no additional hot-dispersion root solve is required. In the Chiu-Chan model, $Z_{\rm R}$ is retained in the real part of the dielectric tensor. The local damping rate is then obtained from a thermally corrected dispersion relation. The plasma dispersion function $Z(\xi_e)$ itself can be evaluated efficiently using standard numerical approximations, but the thermally corrected local root must still be solved repeatedly. The heating model is computationally inexpensive because it retains only $Z_{\rm I}(\xi_e)$ and evaluates the damping term on the cold-dispersion root. The corresponding reduced current-drive efficiency is benchmarked below against the Ehst-Karney parameterization.

### 2.3.2 Regarding current drive model

In this work, the ray-propagation framework is not modified. The ray path, ray weighting, the mapping from real space to flux coordinates, and the flux-surface volume factor are therefore treated as external inputs, as in conventional ray-based modelling. Within this framework, the absorbed-power estimate is controlled by the local imaginary wavenumber along a prescribed propagation path. Its accuracy is assessed in the heating part of this work by benchmarking $k_{\perp i}^{\rm red}$ against the Chiu-Chan damping rate and by evaluating the corresponding single-pass absorption error. For a direct comparison with the Ehst-Karney parameterization, the same dimensional normalization is used. The Ehst-Karney current-drive efficiency is written as[32]:

$$\eta_{\rm EK} = A_{\rm EK}\tilde{\eta}_{\rm EK} \tag{52}$$

with

$$A_{\rm EK} = \frac{38.4\times 10^{18}\cdot T_e}{\ln\Lambda\cdot n_e} \tag{53}$$

In Eq.(53), the electron temperature is given in keV, and the electron density is given in $\rm m^{-3}$. The reduced model is expressed in the same form:

$$\eta_{\rm CD}^{\rm red} = A_{\rm EK}\tilde{\eta}_{\rm LD}^{\rm red} \tag{54}$$

Since the common dimensional factor $A_{\rm EK}$ cancels in the benchmark, the comparison is performed using the dimensionless efficiencies. The Ehst-Karney velocity normalization is

$$w_{\rm EK} = \frac{\omega}{k_{\parallel}\sqrt{T_e / m_e}} \tag{55}$$

which is related to the Landau parameter by:

$$w_{\rm EK} = \sqrt{2}\xi_e \tag{56}$$

Retaining the finite-$w_{\rm EK}$ correction gives the reduced dimensionless efficiency

$$\tilde{\eta}_{\rm LD}^{\rm red} = \left(\frac{3}{w_{\rm EK}Z_{\rm eff}} + \frac{3.83}{Z_{\rm eff}^{0.707}} + \frac{4w_{\rm EK}^2}{5+Z_{\rm eff}}\right)\mathcal{R}_{\rm rel}T_{\rm trap} \tag{57}$$

Two benchmark cases are considered, with increasing physics content in the reference model. In Case A, the reduced model is compared with the Landau-damping branch of the Ehst-Karney parameterization. Following the notation of Ehst-Karney, the reference efficiency is

$$\tilde{\eta}_{\rm caseA}^{\rm ref} = C_{\rm LD}R_{\rm LD}M_{\rm LD}\tilde{\eta}_{\rm LD}^{\rm ref} \tag{58}$$

with

$$\tilde{\eta}_{\mathrm{LD}}^{\mathrm{ref}} = \frac{K_{\mathrm{LD}}}{w_{\mathrm{EK}}} + D_{\mathrm{LD}} + \frac{4w_{\mathrm{EK}}^2}{5+Z_{\mathrm{eff}}} \tag{59}$$

This case is a like-for-like comparison within the same Landau-damping scope. In particular, $\tilde{\eta}_{\mathrm{LD}}^{\mathrm{red}}$ and the Case-A reference share the same LD kernel, $\tilde{\eta}_{\mathrm{LD}}^{\mathrm{ref}}$. Therefore, this benchmark does not re-test the kernel itself. Instead, it isolates the replacement of the Ehst-Karney empirical correction factors, $C_{\mathrm{LD}}R_{\mathrm{LD}}M_{\mathrm{LD}}$, by the more transparent factors $\mathcal{R}_{\mathrm{rel}}T_{\mathrm{trap}}$. This replacement is the central closure made in the reduced current-drive model: a simpler and more interpretable correction structure is used in place of the empirical Ehst-Karney correction, without introducing a new current-drive kernel. The Case-A error is defined as：

$$\delta_{\eta_{\mathrm{A}}} = \frac{\tilde{\eta}_{\mathrm{LD}}^{\mathrm{red}}}{\tilde{\eta}_{\mathrm{LD}}^{\mathrm{ref}}} - 1 \tag{60}$$

In Case B, the reference is extended to include the Alfvén-wave/TTMP branch of the Ehst-Karney model. The corresponding reference efficiency is:

$$\tilde{\eta}_{\mathrm{caseB}}^{\mathrm{ref}} = C_{\mathrm{AW}}R_{\mathrm{AW}}M_{\mathrm{AW}}\tilde{\eta}_{\mathrm{AW}}^{\mathrm{ref}} \tag{61}$$

Since TTMP is omitted in the present reduced model, this case defines a broader reference for the current-drive benchmark. It measures the sensitivity of the predicted efficiency to the damping channel retained in the reference and provides a practical bound on the applicability of the reduced current-drive model. The coefficients $C$, $R$, and $M$, together with the Alfvén-wave kernel $\tilde{\eta}_{\mathrm{AW}}^{\mathrm{ref}}$ are taken directly from the original Ehst-Karney parameterization and are listed in Appendix C.

# 3 Benchmark and Parameter Study

To the authors' best knowledge, the reduced closure derived in section 2.1 is the first single-channel reduced model proposed for helicon-wave damping in RF wave heating and current-drive modelling. In the absence of an established benchmark dataset, its validity must be assessed case by case. We benchmark the model on four representative present-day tokamaks, which together span approximately one order of magnitude in magnetic field, electron density and electron temperature, as listed in Table 1 [17,18,24,43]. These four devices cover distinct regions of the parameter space relevant to tokamak helicon-wave modelling. EAST and HL-3 represent medium-field, medium-temperature operating conditions; DIII-D samples a comparatively low-density fast-wave current-drive regime; and KSTAR provides an independent superconducting long-pulse tokamak case.

For each device, we scan the frequency over $2f_{ci} \le f \le 0.9f_{ce}$ and normalise it by the lower-hybrid frequency, $f/f_{LH}$, to enable direct cross-device comparison. The lower bound is placed above the immediate ion-cyclotron-resonance region to avoid ICRF-like absorption channels that the reduced model does not include. The upper bound is intentionally extended far beyond the engineering-relevant fast-wave operating window, so that the closure can be examined both inside and outside its intended range of applicability. The frequency scan alters the cold-dispersion structure and the relative weight of Landau damping in the full kinetic model**.** This variation, captured by the Landau-dominance ratio $\eta_L$, ultimately governs the validity range of the reduced model.

Table 1. Reference parameters of the four devices used to define the cross-device scan. The quoted values are representative baseline parameters compiled from Refs. [17,18,43–45].

| Device | $B_0(T)$ | $n_{e0}(10^{19}m^{-3})$ | $T_{e0}(KeV)$ | $N_{\parallel 0}$ | $a$(m) | $R_0$(m) | $Z_{\mathrm{eff}}$ |
|---|---|---|---|---|---|---|---|
| EAST | 2.40 | 3.89 | 5.49 | 4.40 | 0.45 | 1.85 | 1.5 |
| HL-3 | 2.20 | 4.00 | 4.00 | 2.88 | 0.65 | 1.78 | 1.5 |
| DIII-D | 1.30 | 2.50 | 1.00 | 2.70 | 0.67 | 1.66 | 1.5 |
| KSTAR | 3.50 | 4.00 | 5.00 | 3.00 | 0.5 | 1.80 | 1.5 |

Although the scan covers the full range $2f_{ci} \lesssim f \lesssim 0.9f_{ce}$, the cold-plasma dispersion itself imposes a structural constraint on where a closure comparison is meaningful. Figure 1(a)-(d) show the propagating $N_\perp$ roots as a function of $f/f_{LH}$ for the four devices. In all cases, only a single helicon-like (low-$n_\perp$) branch propagates for $f \lesssim 0.9 f_{LH}$, while a second high-$N_\perp$ root emerges as the frequency approaches and crosses the lower-hybrid resonance. In this multi-branch region, the cold dispersion alone cannot identify which root a ray-tracing code should follow: the two branches couple near the mode-conversion layer, and the error becomes inseparable from the ambiguity of branch selection.

To verify that this transition is generic rather than configuration-specific, Figure 1(e) reports, for each device, the fraction of randomly sampled equilibrium points within its operating envelope at which more than one propagating $N_\perp$ root exists, plotted against $f/f_{LH}$. All four devices show essentially zero multi-branch occurrence below $f \approx 0.9f_{LH}$ and a sharp rise immediately above it. The transition frequency is consistent across devices despite the order-of-magnitude differences in their nominal magnetic field, density and temperature. The $0.9f_{LH}$ boundary is therefore a property of the cold dispersion itself, not of the closure under test, and represents a constraint orthogonal to the closure's intrinsic accuracy. On this basis, the frequency axis is partitioned into two regions, indicated by the green and red shading in Figure 1(a)-(e):

(i). Sub-LH window ($2f_{ci} \lesssim f \lesssim 0.9f_{LH}$, green): single-branch helicon wave propagation, well-posed for benchmarking. All quantitative validity statements in section 4 are restricted to this window.

(ii). Branch-ambiguous extension ($f \gtrsim 0.9f_{LH}$, red): branch-ambiguous or topology-changing region. Scan points in this region are retained in the dataset and reported as out-of-domain reference values where appropriate, but are excluded from the validity domain established in this work. The branch behaviour in this extended region is further examined using branch-locked all-hot computations in Appendix A.

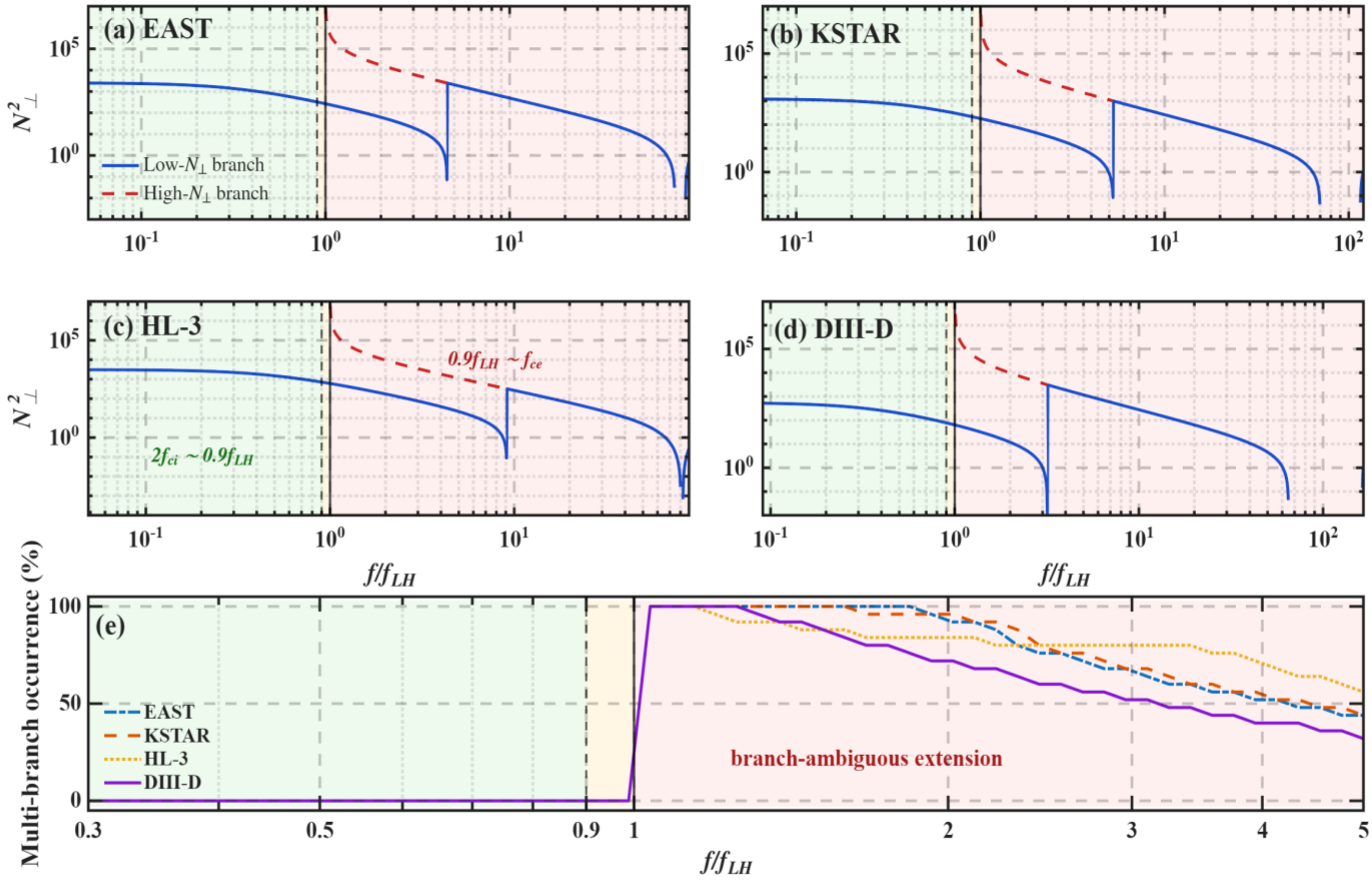

Figure 1. Cold-dispersion branch structure and frequency-domain partition: (a)-(d) EAST, KSTAR, HL-3 and DIII-D; (e) shows the corresponding multi-branch occurrence probability, evaluated from each case local plasma parameters within each device.

To systematically characterise the error of the reduced model across parameter space, we scan each of the four devices along two key directions within the sub-LH window: electron temperature $T_e$ and parallel refractive index $n_{\parallel}$. The scan ranges are:

$$\begin{aligned} T_e &= T_{e0} \times a_0 \qquad a_0 \in [0.5,\ 2.5] \\ N_{\parallel} &= N_{\parallel 0} \times a_1 \qquad a_1 \in [0.5,\ 1.5] \end{aligned} \tag{62}$$

where $T_{e0}$ and $N_{\parallel 0}$ denote the device-specific nominal values listed in Table 1. The temperature interval is chosen to cover substantial variations around the nominal thermal state, from lower-temperature conditions to hotter core-like regimes. The $N_{\parallel}$ interval is chosen to represent a broad but still experimentally relevant range of launched parallel refractive indices, while avoiding unrealistically small or excessively large values. Combined across the four devices, $T_e$ spans from 0.50 to 13.725 keV and the $N_{\parallel}$ spans from 1.35 to 6.6, covering the main parameter range relevant to experimental helicon wave heating and current drive. We take twenty equally spaced sampling points along each axis, yielding 400 cases per device and 1600 cases in total. For each case, the frequency is further scanned over 1000 equally spaced points within the sub-LH window.

# 4 Validation and Correction of Wave Heating Model

Although both the Chiu-Chan benchmark and the reduced model admit form evaluation at a prescribed $(T_e, N_{\parallel}, f)$, their algebraic structures are different. In the Chiu-Chan formulation, the damping correction to $k_{\perp i}$ is evaluated through the wave-damping factor $G$, which depends on the full Fried-Conte plasma dispersion function $Z(\xi_e)$ and on the composite quantities entering the kinetic dispersion relation. By contrast, the reduced model retains only the leading Landau-pole contribution, proportional to $\sqrt{\pi}\xi_e \exp(-\xi_e^2)$, and obtains $N_{\perp}^2$ from a single quadratic equation tied to the cold-helicon branch. This shorter evaluation path enables a systematic scan over approximately $1.6\times10^6$ points on a single workstation. The key question is whether this simplification preserves sufficient accuracy across the relevant parameter space. The following subsection therefore quantifies the accuracy loss and identifies the parameter regimes in which the reduced model remains reliable.

**4.1 Accuracy and error decomposition of the reduced model**

The nominal-point frequency scans reveal two distinct error behaviours among the four devices. Figure 2 shows the frequency dependence of the reduced-model error, $\delta_k^a$, and the Landau-dominance ratio $\eta_L$ for the four devices at their nominal operating points. The shaded green band denotes the strict low-error region, defined here as errors below 5%, while the shaded yellow band denotes a secondary low-error region, defined as errors below 20%. As shown in Figure 2(a) (b) and 2(c), EAST, KSTAR and HL-3 show a decrease in $\delta_k^a$ with increasing frequency. As a result, these devices exhibit a low-error interval. In these cases, the reduced model does not suffer from a one-sided systematic bias: it overestimates $|k_{\perp i}|$ at lower $f/f_{\mathrm{LH}}$ and underestimates it at higher $f/f_{\mathrm{LH}}$, with the sign reversal occurring inside the low-error band. Importantly, the low-error window is reached before $\eta_L$ saturates, indicating that accurate prediction does not require Landau damping to be the overwhelmingly dominant absorption channel. As shown in Figure 2(d), DIII-D exhibits the opposite behaviours. In this case, $\eta_L$ rapidly approaches unity at $f/f_{\mathrm{LH}} \simeq 0.05-0.2$ and remains saturated over most of the scanned window. However, $\delta_k^a$ does not decrease accordingly; instead, it remains at a finite plateau of about 0.36. Thus, even when the damping is almost entirely Landau-dominated, the reduced model does not necessarily converge to the Chiu-Chan benchmark. This behaviour is consistent with the error decomposition in Eq.(50). Once the absorption becomes almost entirely Landau dominated, the non-Landau truncation error is already small. The remaining error is then governed mainly by the residual mismatch within the retained Landau channel, represented by the factor $(\rho-1)$. Therefore, the large errors observed for DIII-D cannot be attributed simply to the omission of non-Landau absorption channels. Instead, they indicate that the leading Landau-pole surrogate does not fully reproduce the Landau-channel response of the Chiu-Chan model in these parameter regimes.

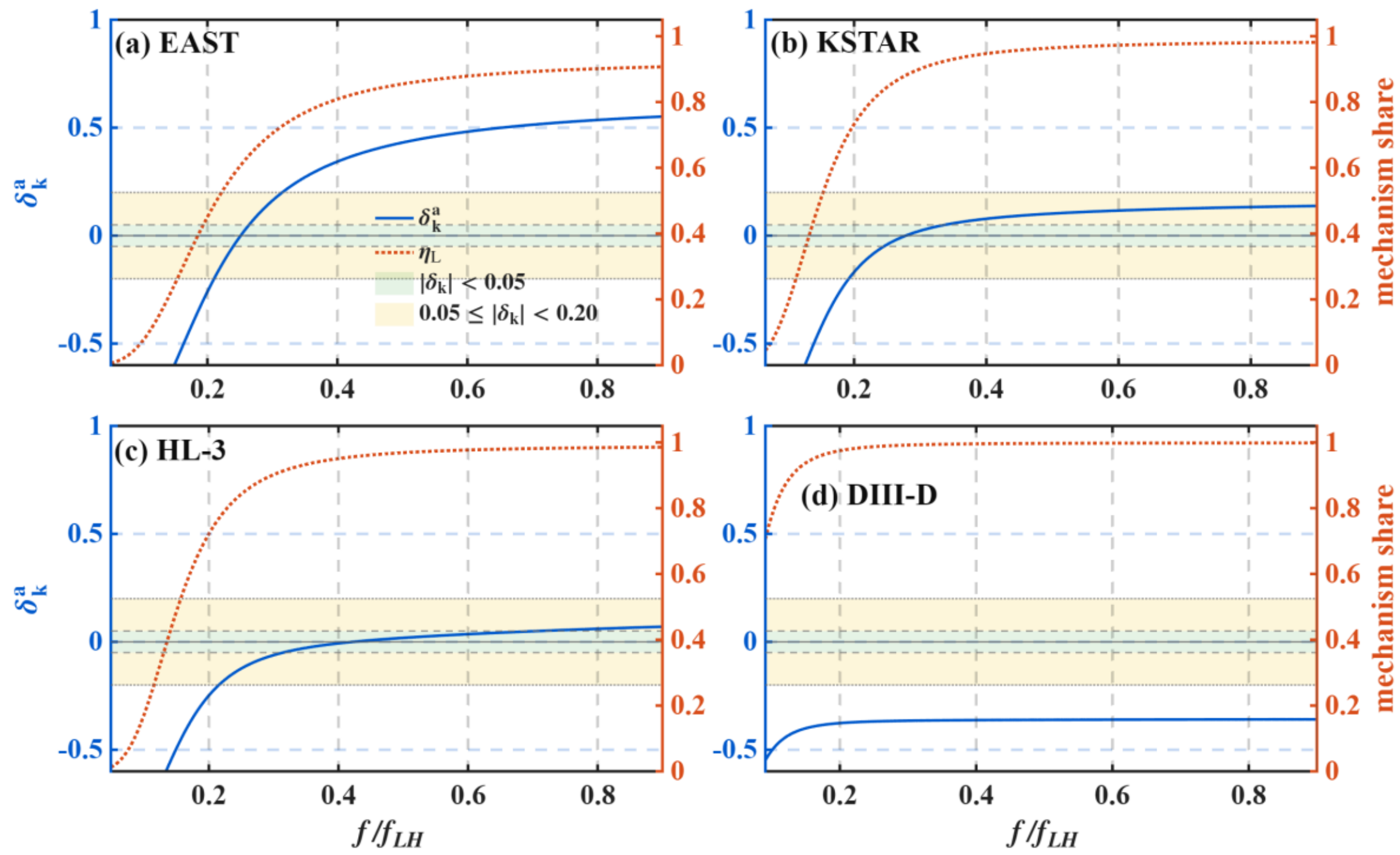

Figure 2. Evolution of $\delta_k^a$ and $\eta_L$ versus frequency for four devices under typical operating conditions; (a) EAST (b) KSTAR (c) HL-3 (d) DIII-D.

The low-error region occupies only a limited part of the scanned parameter space. Figure 3 shows the coverage of the low-error region, defined as $|\delta_k^a| \le 0.2$, over the scanned parameter space. The coverage denotes the percentage of cases within each bin that satisfy this error criterion. Figure 3(a) shows the aggregated coverage in the $(f/f_{\mathrm{LH}}, \log_{10}\xi_e)$ plane for all four devices, whereas Figure 3(b)-(e) show the corresponding device-resolved coverage in the $(N_\parallel, \log_{10}\beta_e)$ plane for EAST, KSTAR, HL-3 and DIII-D, respectively. The aggregated map in Figure 3(a) shows that high coverage is confined to a limited region of parameter space. The main low-error band is centred near $\log_{10}\xi_e \simeq 0-0.6$ and $f/f_{\mathrm{LH}} \simeq 0.15-0.5$. Within this band, the coverage can exceed 80%, whereas it decreases rapidly outside it. This indicates that the validity of the reduced model is not controlled by frequency alone, but is strongly organized by the Landau parameter $\xi_e$ within an appropriate sub-LH frequency interval.

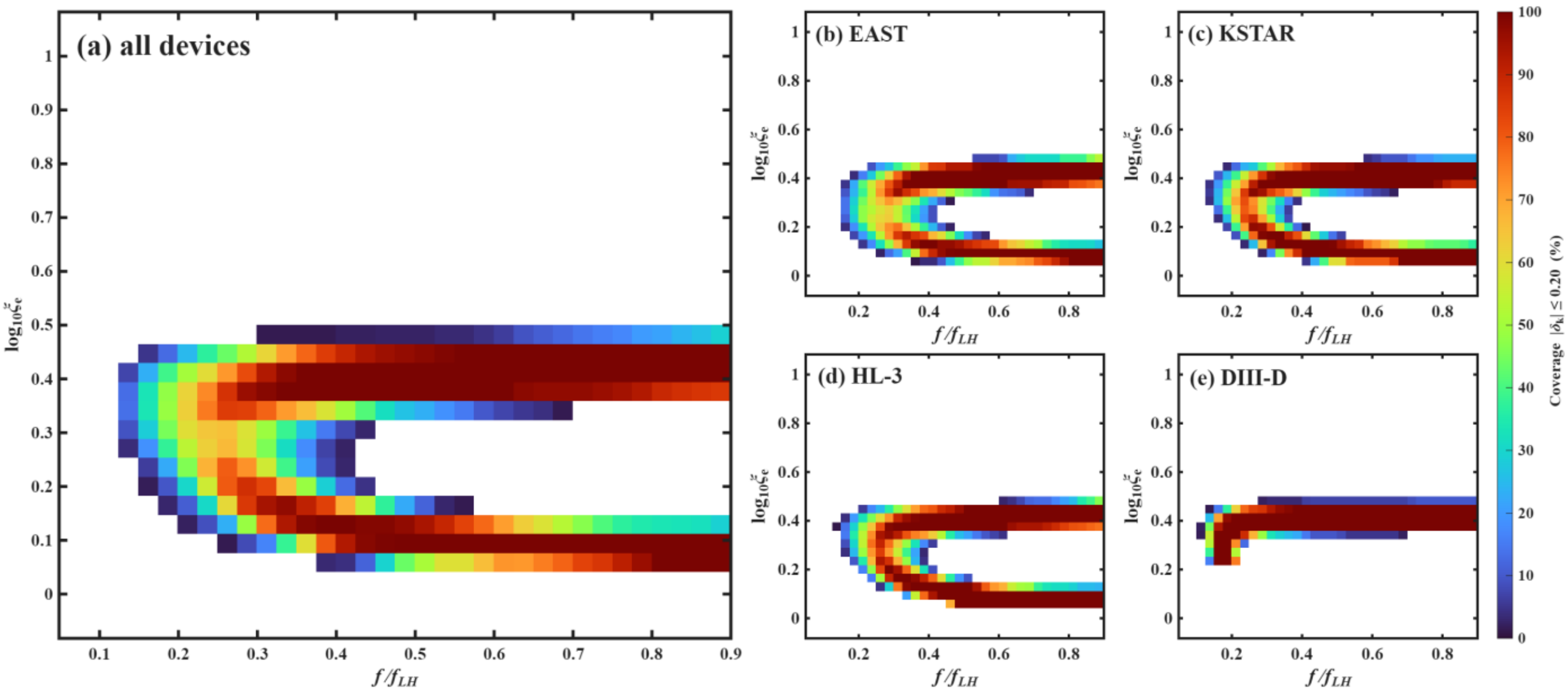

Figure 3. Low error coverage of the reduced model over the scanned parameter space: (a) shows the aggregated map in the $(f/f_{\mathrm{LH}}, \log_{10}\xi_e)$ plane for all devices; (b) - (e) show the device-resolved maps in the $(N_\parallel, \log_{10}\beta_e)$ plane for EAST, KSTAR, HL-3 and DIII-D.

The device-resolved maps in Figure 3(b)-(e) further support this interpretation. For each device, the high-coverage region forms a diagonal band in the $(N_\parallel, \log_{10}\beta_e)$ plane. This structure reflects the dependence of $\xi_e$ on $N_\parallel$ and $T_e$ through $\xi_e = \omega/(k_\parallel v_{te}) = c/(N_\parallel v_{te})$ together with the variation of $\beta_e$ with $T_e$ at fixed device density and magnetic field. Consequently, an increase in $N_\parallel$ can be compensated by a decrease in $T_e$ or $\beta_e$, and vice versa, so that $\xi_e$ remains in the low-error interval. The position and width of the diagonal band differ between devices because $\beta_e$ also contains the device-dependent factors $n_e$ and $B_0$. Taken together, Figure 3 shows that the low-error region is not tied to any single nominal operating point. Instead, it is organised by a joint constraint: the wave must remain within a suitable sub-LH frequency window, and $\xi_e$ must fall in the range where the reduced Landau-only approximation remains accurate. This provides the empirical basis for adopting $Z_R(\xi_e)$ as the natural variable in the empirical correction developed in the following subsection.

For each scanned parameter point, the frequency that gives the minimum error within the prescribed scan window is identified as $f_* = \arg\min_f |\delta_k^a|$. The corresponding value of $\delta_k^a$ is then used to characterise the best attainable accuracy of the reduced model at that point. Figure 4 projects these results onto the $(\xi_e, \beta_e)$ plane, with $\xi_e$ is evaluated at $f_*$. The cases are divided into three error tiers: $|\delta_k^a| < 0.05$ for the low-error region, $0.05 \le |\delta_k^a| < 0.2$ for the transition region, and $|\delta_k^a| \ge 0.2$ for the breakdown region. Figure 4 shows that the three error tiers are clearly organized in the $(\xi_e, \beta_e)$ plane. Low-error cases concentrate at $0.5 \le \xi_e \lesssim 3$. Outside this region, the reduced model progressively moves from the low-error tier into the transition and breakdown tiers, particularly as $\xi_e$ increases or decreases. The four devices populate this plane very differently. As shown in Figure 4(b), (c) and (d), EAST and HL-3 contain broad and continuous low-error regions. By contrast, as shown in Figure 4(e) DIII-D is dominated by breakdown-tier cases. The validity of the reduced model is determined not by the device itself, but by where its operating envelope lies in the $(\xi_e, \beta_e)$ plane.

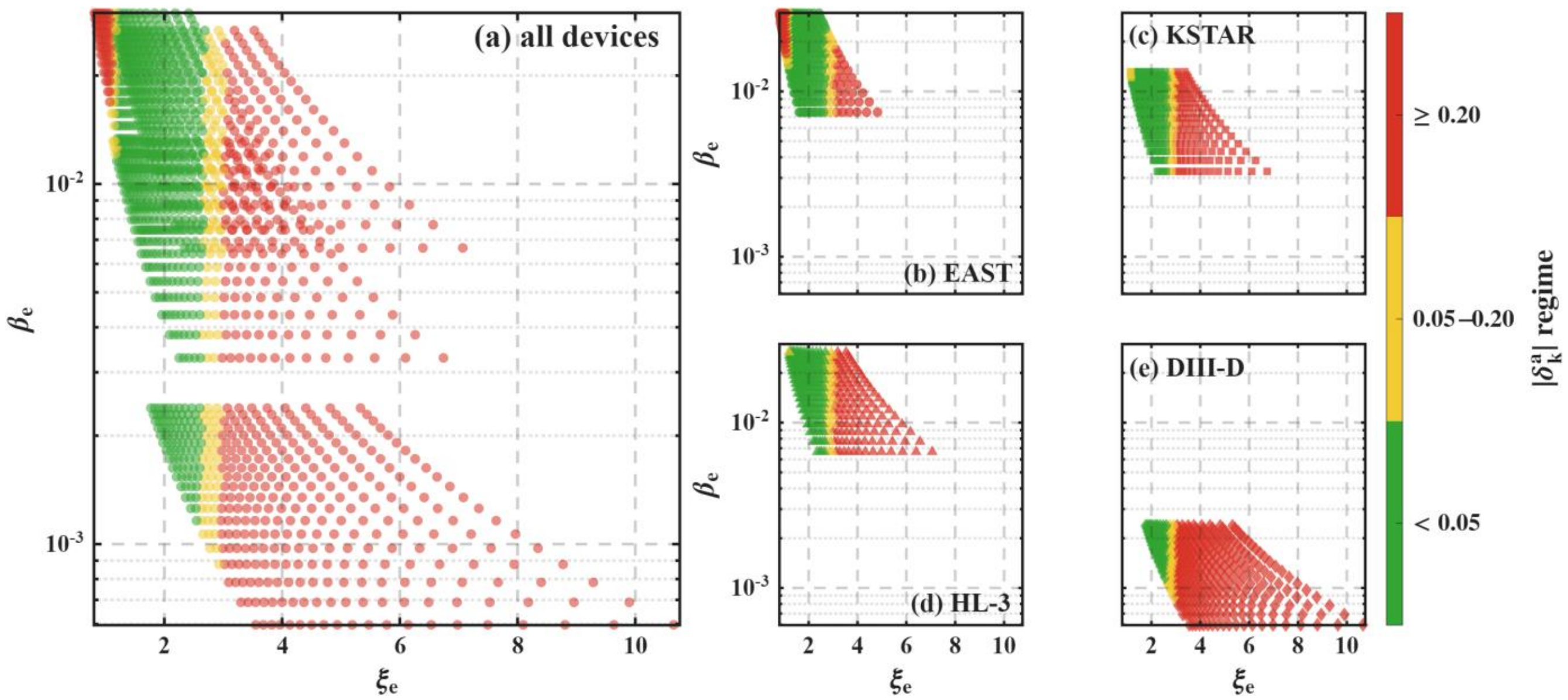

Figure 4. Distribution of $\delta_k^a$ in the $(\xi_e, \beta_e)$ plane from the parameter scan of four devices: (a) All Devices; (b) EAST; (c) KSTAR; (d) HL-3; (e) DIII-D.

These two coordinates impose complementary constraints on the validity of the reduced model. The Landau parameter, $\xi_e$, measures how close the parallel phase velocity of the wave is to the electron thermal velocity. Efficient Landau interaction occurs when these two velocities are comparable. If the parallel phase velocity is too high, only a small fraction of electrons can resonate with the wave, and the retained Landau response becomes weak. The electron beta, $\beta_e$, measures how strongly the electron thermal response modifies the cold-plasma background. If $\beta_e$ is too small, this thermal correction becomes dynamically negligible. Accurate reduced-model predictions therefore require both a suitable resonant-electron population and a sufficiently strong thermal response. The $(\xi_e, \beta_e)$ plane thus defines a joint validity boundary that cannot be determined from either coordinate alone.

To identify the origin of the reduced-model error, we use the decomposition shown as Eq.(50). The first term $-(1-\eta_L)$ represents the truncation error from neglecting non-Landau damping channels, whereas the second term $\eta_L(\rho-1)$ represents the residual approximation error within the retained Landau contribution. Figure 5 plots the scan cases for each device in the $(\eta_L, \delta_k^a)$ plane, together with the two decomposed branches.

The error decomposition explains why Landau dominance alone does not guarantee a small total error. Figure 5 shows a qualitatively consistent structure across the four devices. The dashed black curve, $-(1-\eta_L)$, increases monotonically toward zero as $\eta_L$ approaches unity, indicating that the truncation error is progressively suppressed when the system enters the Landau-dominated regime. However, the residual error within the retained Landau channel, $\eta_L(\rho-1)$, does not vanish automatically with increasing $\eta_L$. Instead, its magnitude is controlled by the residual factor $\rho-1$, which measures how well the reduced Landau-pole surrogate reproduces the corresponding Landau-channel response of the Chiu-Chan model. As the result, the total error $\delta_k^a$ can become small either because both decomposed terms are individually small, or because two finite contributions of opposite sign partially cancel.

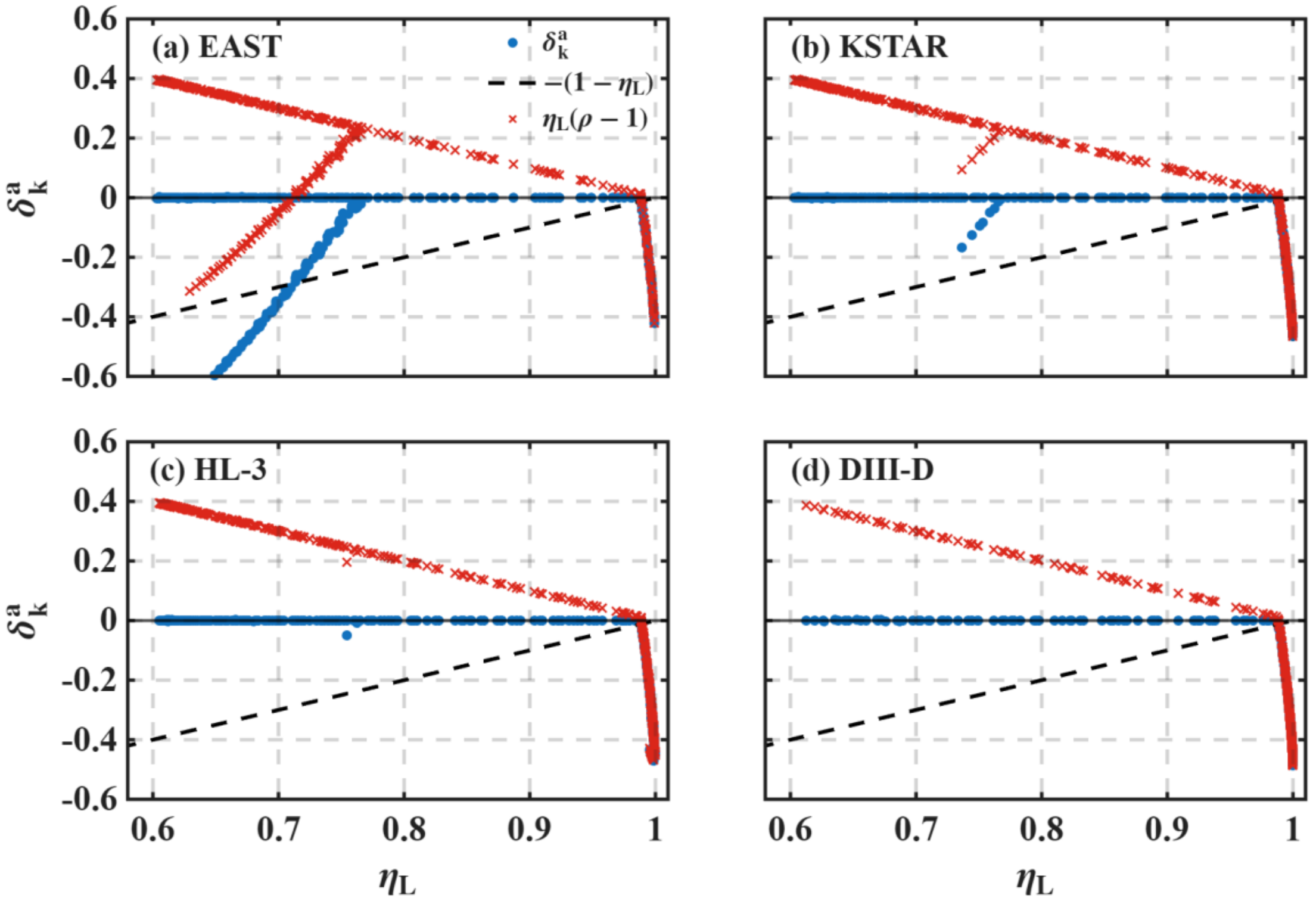

Figure 5 Error decomposition of the cases from the four-device scan in the $(\eta_L, \delta_k^a)$ plane, Dots: total error $\delta_k^a$; black dashed line: truncation branch $-(1-\eta_L)$; red crosses: approximation branch $\eta_L(\rho-1)$; (a) EAST; (b) KSTAR; (c) HL-3; (d) DIII-D.

This decomposition clarifies the role of $\eta_L$. A large $\eta_L$ suppresses the error associated with omitted non-Landau channels, but it does not by itself guarantee high accuracy of the reduced model. In the Landau-dominated regime, the remaining error is governed by the behaviour of $\rho-1$. The internal accuracy of the reduced model therefore reduces to the question of how $\rho$ varies with the kinetic state. The following subsection shows that, within a restricted parameter window, $\rho$ exhibits a pronounced cross-device collapse when plotted against $Z_R(\xi_e)$, providing the basis for the empirical correction.

**4.2 Empirical correction and validation**

In this work, $Z_R(\xi_e) \equiv \Re[Z(\xi_e)]$ is defined as the real part of the Fried-Conte plasma dispersion function $Z(\xi_e)$. For real $\xi_e$, it is given by

$$Z_R(\xi_e) = \Re[Z(\xi_e)] = -2e^{-\xi_e^2}\int_0^{\xi_e} e^{t^2}\,dt \tag{63}$$

$Z_R(\xi_e)$ serves as a dimensionless measure of the Hermitian dispersive feedback associated with the electron parallel thermal response. The real part of this thermal response nearly vanishes when the parallel phase velocity lies far below the electron thermal velocity. It becomes strongly negative once the two velocities are comparable. In this regime the real part substantially modifies wave propagation and the dispersion relation. It returns toward zero when the phase velocity far exceeds the thermal velocity, because the wave then outpaces the thermal electrons and the response weakens.

The real and imaginary parts of the Fried-Conte dispersion function, $Z_R$ and $Z_I$, correspond to the Hermitian and anti-Hermitian parts of the same electron resonant response, respectively. For real $\xi_e$, $Z_I(\xi_e)=\sqrt{\pi}e^{-\xi_e^2}$ gives the anti-Hermitian response associated with the Landau resonance. By contrast, $Z_R(\xi_e)$ gives the accompanying Hermitian thermal dispersive feedback. This feedback enters the real part of the dielectric tensor and modifies the associated dispersion structure. The reduced model retains the absorptive imaginary part produced by $Z_I$, but still uses the cold-plasma dielectric tensor to solve for the real root and evaluate the wave-response combinations. The reduced model therefore retains the anti-Hermitian part of Landau damping while neglecting the feedback of $Z_R$, which enters the real part of the hot dielectric tensor and modifies the propagation structure. In contrast, the Chiu-Chan model retains $Z_R(\xi_e)$ through the composite quantities defined in Eqs.(41) and (42).

Figure 6(a) quantifies the reduced-model error by showing the dependence of $\rho$ on $Z_R$. Within $-1\le Z_R(\xi_e)\le -0.2$, the $\rho$ values from the four devices approximately collapse onto a single-valued function of $Z_R$. This collapse indicates that, within this restricted window, the residual error of the reduced model relative to the Chiu-Chan benchmark is governed primarily by the Hermitian thermal feedback represented by $Z_R$. On this basis, $\rho(Z_R)$ is parameterized by an empirical response expansion satisfying the normalization constraint $\rho(0)=1$, as shown in Figure 6(b):

$$\begin{aligned}&\rho_{\text{fit}}(Z_R)=1+\sum_{j=1}^{5}a_j Z_R^j\\&a_1=4.7487,\quad a_2=13.6959,\quad a_3=0.0670,\\&a_4=-18.2025,\quad a_5=-9.6866\end{aligned}\tag{64}$$

The fit gives $R^2=0.9986$ over $-1\le Z_R(\xi_e)\le -0.2$, showing that $Z_R$ captures the dominant statistical variation of $\rho$. Substituting Eq.(64) into the error decomposition in Eq.(50) yields the $Z_R$ based prediction of the residual error:

$$\hat{\delta}_k^a=\eta_L\,\rho_{\text{fit}}(Z_R)-1\tag{65}$$

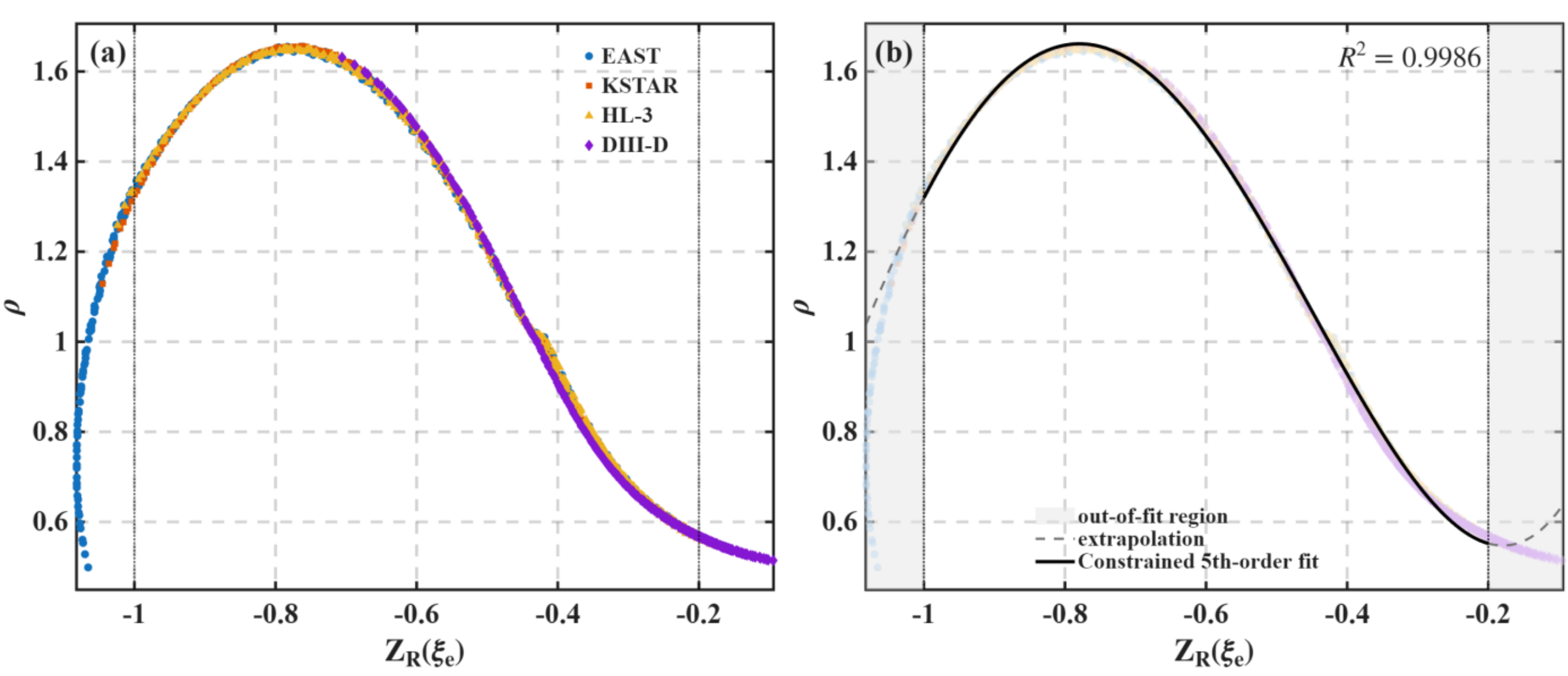

Figure 6 Relation between the closure-quality factor $\rho$ and $Z_R(\xi_e)$ distribution: (a) full $\rho$ - $Z_R$ distribution for the four devices; (b) Collapse of $\rho$ in the fitting window and the constrained fifth-order empirical response expansion.

The predictive performance of $\rho_{\text{fit}}(Z_R)$ is assessed by comparing $\hat{\delta}_k^a$ from Eq.(65) with the directly evaluated error $\delta_k^a$ at each scan point. As shown in Figure 7(a), most samples within the fitting window have small prediction residuals, indicating that the difference between the predicted and directly evaluated errors remains close to zero. The median absolute prediction error is 0.0073, and the 90th-percentile error (P90) is 0.0185. Throughout this work, P90 denotes the 90th percentile of a distribution: 90% of the samples fall below this value. These statistics show that $\rho_{\text{fit}}(Z_R)$ captures the dominant variation of the reduced-model error across the four devices within this window. Outside the fitting window, the prediction residuals become much larger and show a structured distribution, as shown in Figure 7(b). The median absolute prediction error increases to 0.0182, and the P90 error increases to 0.1813. The increase is not only a broadening of the residual distribution; the residuals also show a structured pattern. This indicates that $\rho_{\text{fit}}(Z_R)$ no longer captures all of the relevant variation in the reduced-model error. The $Z_R$-based empirical relation should therefore not be extrapolated as a globally valid correction. Its validity is mainly confined to the resonant window identified above. Once the system leaves this window, the residual error is no longer governed solely by the Hermitian thermal feedback represented by $Z_R$. Other factors, including the frequency range, branch structure, and the relative weight of the cold-plasma propagation terms, may then become important.

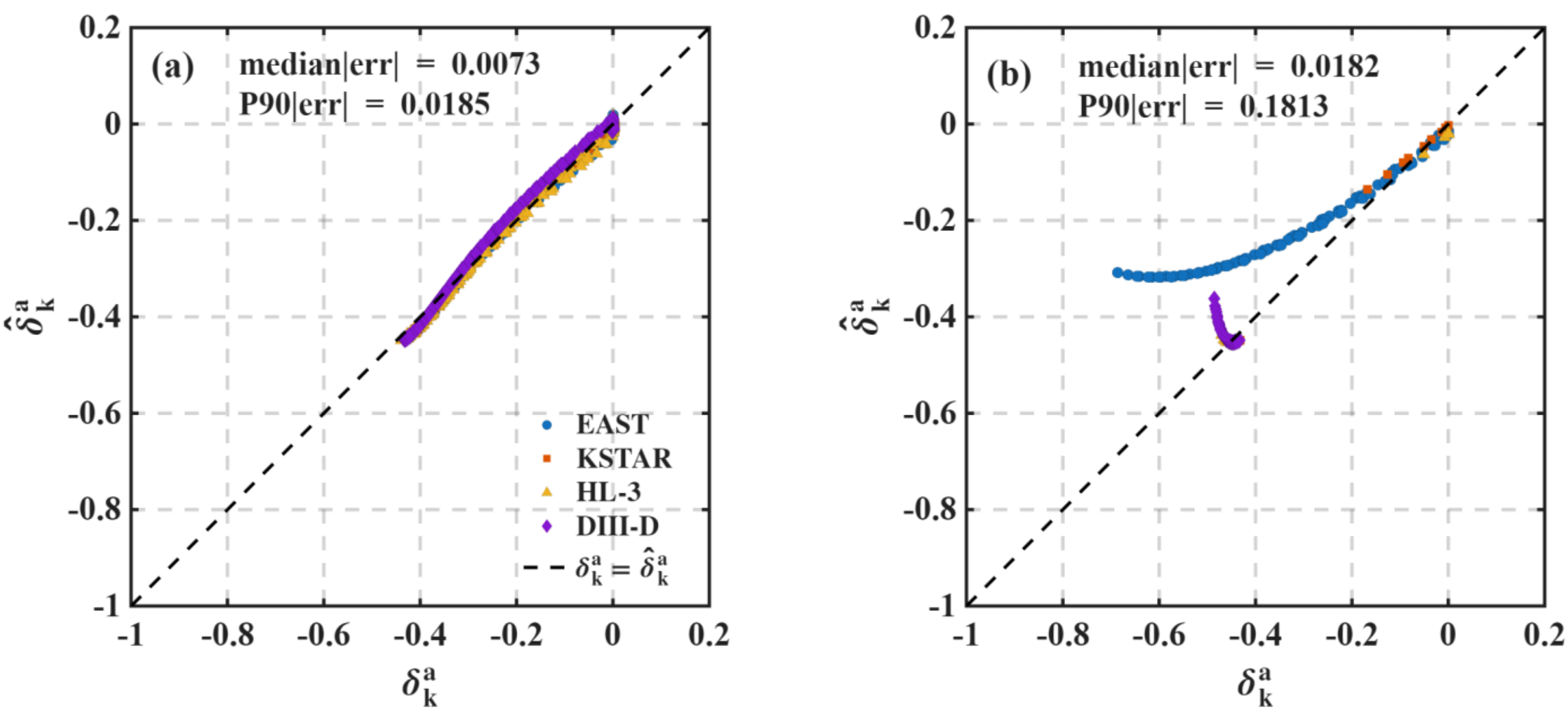

Figure 7. Validation of the error prediction based on the fitted $\hat{\rho}(Z_R)$: (a) inside the fitting window; (b) outside the fitting window.

Taken together, Figure 6 and Figure 7 show that $\rho_{\text{fit}}(Z_R)$ provides an approximately device-independent empirical correction. Within the fitting window, this correction improves the quantitative accuracy of the reduced model while preserving its computational efficiency. However, the error prediction in Eq.(65) still contains $\eta_L$, which is computed from the Chiu-Chan composite quantities and not available to the reduced model alone. This would defeat the purpose of the reduced model in the first place. A self-contained correction is therefore needed.

The channel-truncation error in Eq.(50) measures the contribution deliberately dropped when the non-Landau channels were removed. Within the reduced model, the ELD is the only retained damping channel. Therefore, only the intra-channel approximation error can be corrected internally. The reduced-model output is therefore rescaled directly by $\rho_{\text{fit}}(Z_R)$. The validity-domain restriction keeps the truncation contribution small in the region where the correction is applied. This gives the self-contained correction:

$$k_{\perp i}^{\text{corr}} = \frac{k_{\perp i}^{\text{red}}}{\rho_{\text{fit}}(Z_R(\xi_e))} \tag{66}$$

with the corresponding residual error:

$$\delta_k^{\mathrm{corr}} = \frac{1+\delta_k^a}{\rho_{\mathrm{fit}}(Z_R(\xi_e))} - 1 \tag{67}$$

The empirical correction mainly improves the quantitative accuracy inside the previously identified resonance-controlled validity region. Figure 8 shows the low-error coverage after applying the correction in Eq.(66). Compared with Figure 3, the low-error region is defined by a stricter criterion, $|\delta_k^{\mathrm{corr}}| \le 0.05$. The two red dashed lines in Figure 8(a) mark the fitting window expressed in terms of the corresponding $\log_{10}\xi_e$ range. Compared with the uncorrected coverage map, the corrected model substantially expands the low-error region. In Figure 8(a), high coverage is concentrated within the $Z_R$ fitting window and spans a broad frequency interval, $f/f_{\mathrm{LH}} \simeq 0.2-0.9$. Within this region, the coverage frequently exceeds 80%, indicating that the empirical response $\rho_{\mathrm{fit}}(Z_R)$ removes the dominant residual bias of the reduced model across main resonant band. The device-resolved maps in Figure 8(b)-(e) further show that the correction improves the low-error coverage across all four devices, although the accessible high-coverage region remains device dependent. The corrected low-error region still appears as a diagonal band, reflecting the same $\xi_e$ constraint identified in the uncorrected model. However, the band becomes broader and reaches higher coverage values after applying $\rho_{\mathrm{fit}}(Z_R)$. This indicates that the correction does not remove the underlying validity boundary set by the Landau-resonant parameter range. Instead, it substantially improves the quantitative accuracy within that boundary.

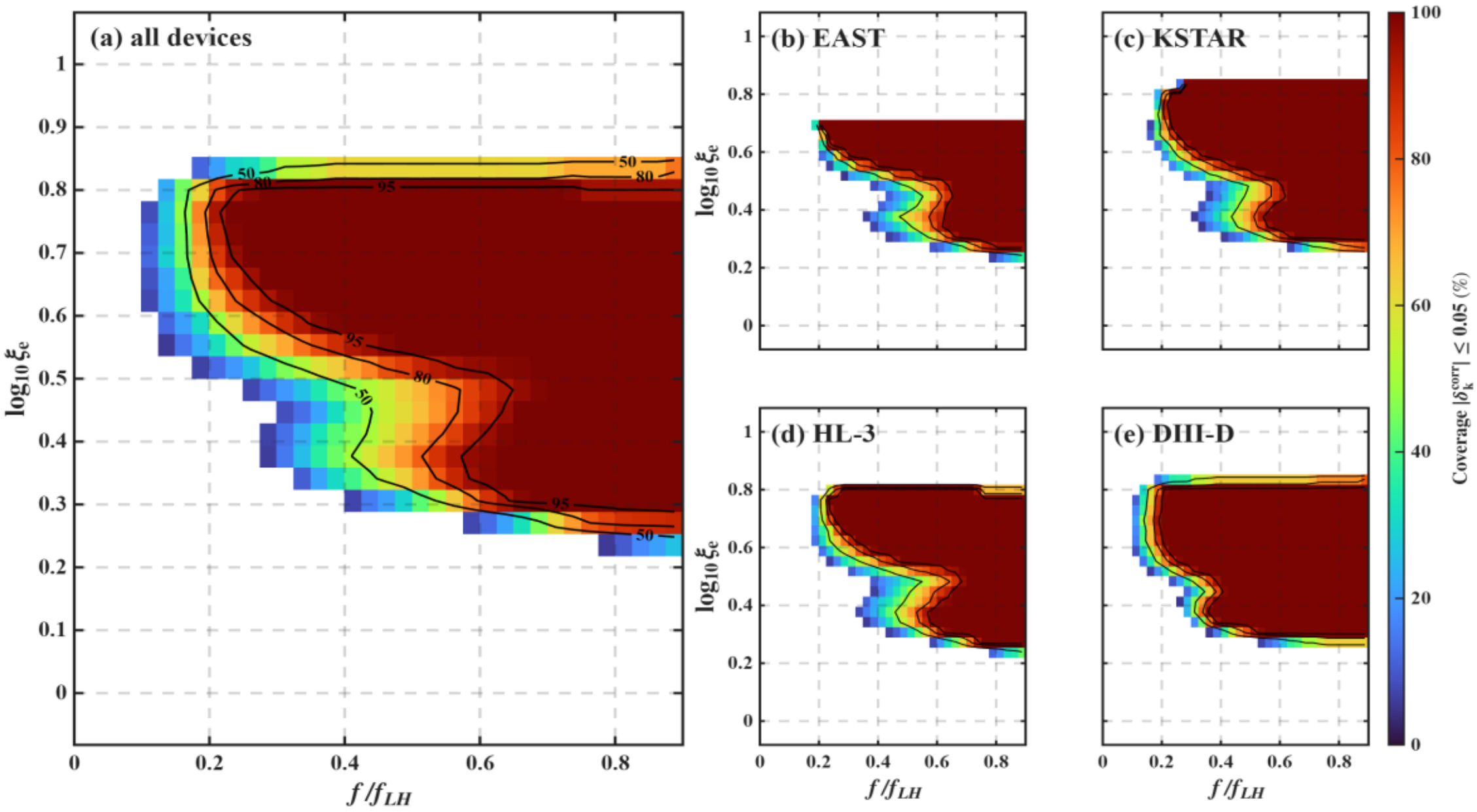

Figure 8. Low-error coverage of the corrected reduced model: (a) all devices; (b) EAST; (c) KSTAR; (d) HL-3; (e) DIII-D.

To test whether $\rho_{\mathrm{fit}}(Z_R)$ generalizes beyond the four calibration devices, two additional tokamak parameter sets are used as hold-out validation cases, referred to here as the ITER-like and BEST-like cases [44,45]. These labels denote scans based on representative theoretical parameters for ITER and BEST (Burning-plasma Experimental Superconducting Tokamak), rather than measurements from experimental profiles. The corresponding parameter values are listed in Table 2.

Table 2 Representative theoretical parameters and expected helicon-wave operating frequencies for ITER and BEST[44,45].

| Device | ITER-like | BEST-like |
|---|---|---|
| $B_0(T)$ | 5.30 | 6.15 |
| $n_{e0}$ ($10^{20}\,\mathrm{m}^{-3}$) | 1.00 | 1.00 |
| $T_{e0}(keV)$ | 8.80 | 10.00 |
| $N_{\parallel 0}$ | 3.00 | 3.00 |
| Frequency window (MHz) | 36-887 | 55-134 |

The hold-out test first checks whether the fitted $\rho_{\text{fit}}(Z_R)$ relation is preserved outside the four calibration devices. Figure 9 shows the distribution of the ITER-like and BEST-like cases in the $(Z_R, \rho)$ plane. Within the fitting interval, both parameter sets follow the $\rho_{\text{fit}}(Z_R)$ curve trained on the four calibration devices. This agreement indicates that the empirical relation remains valid under high-field, high-density operating conditions. Quantitatively, the ITER-like cases yield a median residual of 0.0074 relative to $\rho_{\text{fit}}(Z_R)$, and the BEST-like cases 0.0071. For comparison, the median residual on the four calibration devices is 0.014. The hold-out residuals for the ITER-like and BEST-like cases therefore have the same order of magnitude as the in-sample residuals. This supports the generalisation of the fit beyond the calibration set to these two parameter regimes.

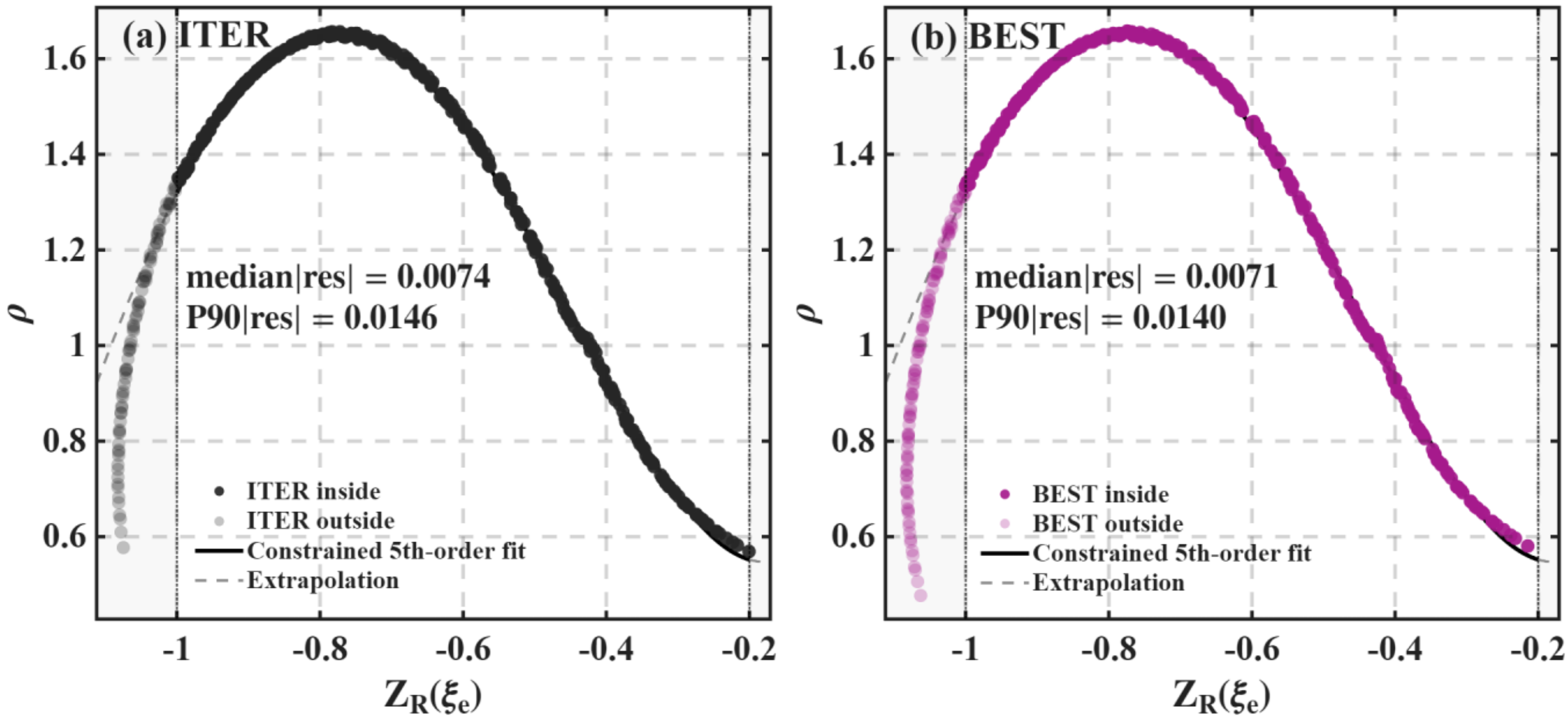

Figure 9. Extrapolation test of the empirical closure using ITER-like and BEST-like hold-out cases, the black solid line denotes the fifth-order polynomial fit trained on the four calibration devices: (a) ITER-like; (b) BEST-like.

A second hold-out test examines whether the same $\rho_{\text{fit}}(Z_R)$ relation can predict the reduced-model error in the ITER-like and BEST-like parameter regimes. Figure 10 compares the predicted error, $\hat{\delta}_k^a$, with the directly scanned error, $\delta_k^a$. The ITER-like and BEST-like cases cluster along the diagonal over most of their range, showing that the relation calibrated on the four reference devices reproduces the total error in these new parameter regimes with comparable accuracy. Taken together, the two hold-out tests support the transfer of the $\rho_{\text{fit}}(Z_R)$ correction to the tested high-field and high-density conditions. Within the restricted $Z_R$ window, the variation of $\rho$ is therefore organized primarily by $Z_R(\xi_e)$, and the resulting $\rho_{\text{fit}}(Z_R)$ provides an approximately device-independent empirical correction over the calibration and hold-out ranges considered here.

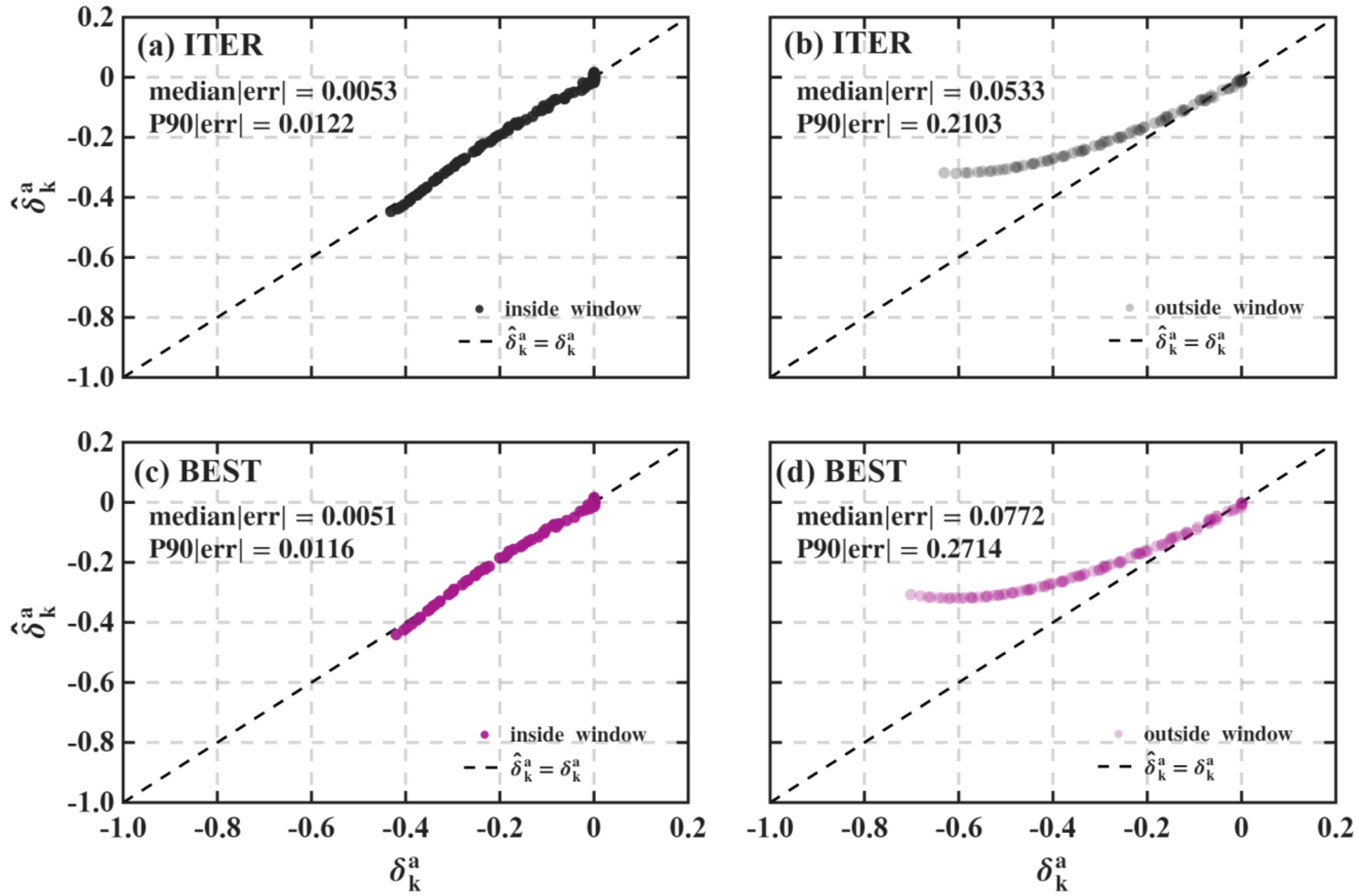

Figure 10 Prediction of $\delta_k^a$ for (a) and (c) ITER-like; (b) and (d) BEST-like.

**4.3 Failure boundary above the lower-hybrid frequency**

The empirical correction developed above was developed and validated within the sub-LH window. Within this window, the corrected reduced model reproduces the reference damping accurately. Above the lower-hybrid frequency, however, the cold-plasma branch structure, wave polarisation and relative contributions of different absorption channels can change substantially. It is necessary to assess whether the same correction remains applicable outside the window.

The correction only gives a bounded recovery outside the sub-LH window. Figure 11 compares the low-error coverage of the reduced model before and after correction outside the sub-LH window. The low-error criterion is the same as that used in Figure 8. Figure 11(a) shows that the uncorrected reduced model has very limited validity in this extended frequency range. The low-error coverage remains close to zero over most of the parameter space. After correction, the low-error region partially recovers, as shown in Figure 11(b). This recovery is confined to a finite domain. The high-coverage region lies mainly within the $Z_R$ fitting interval bounded by the two red dashed lines, and the coverage drops sharply outside this interval. Figure 11(b) also indicates an additional limitation of the correction. Even within the nominal $Z_R$ fitting interval, the low-error coverage decreases as $f / f_{\mathrm{LH}}$ increases and nearly vanishes at the highest frequencies considered, $f / f_{\mathrm{LH}} \approx 25$. This behaviour shows that the validity of the correction is not controlled by the $Z_R$ bounds alone. A frequency-dependent constraint also appears, probably associated with changes in branch structure and polarisation response in the extended frequency regime.

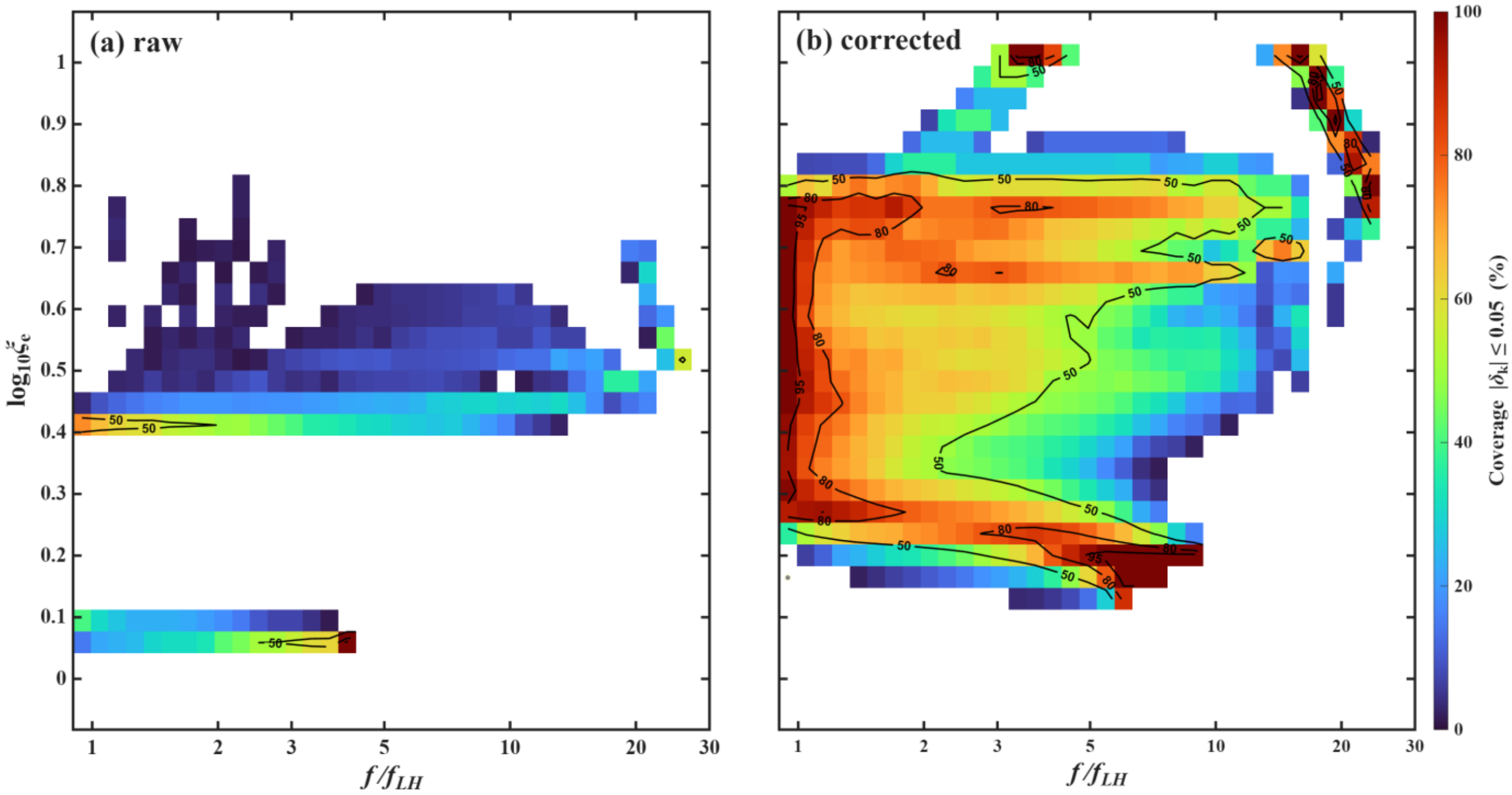

Figure 11. Low-error coverage of the reduced model outside the sub-LH frequency window (a) before and (b) after applying the empirical correction.

To test whether the residual error outside the sub-LH window is merely a fitting limitation, a more flexible two-dimensional empirical fit is introduced and compared with the original one-dimensional correction $\rho_{\text{fit}}(Z_R)$. The two-dimensional fit adds an explicit dependence on the normalised frequency, $x = \log_{10}(f / f_{\text{LH}})$, and is written as

$$\rho_{\text{fit}}^{\text{2D}}(Z_R, x) = 1 + \sum_{j=1}^{5} Z_R^j \sum_{k=0}^{2} a_{j,k} x^k \tag{68}$$

If the residual error outside the sub-LH window mainly reflects missing frequency dependence in the one-dimensional fit, adding $f / f_{\text{LH}}$ should substantially reduce that error. Conversely, if the two-dimensional fit does not materially reduce the error, the breakdown outside the sub-LH window cannot be attributed simply to insufficient flexibility of the empirical fit. It would instead indicate that the reduced model is missing part of the relevant physical structure. Figure 12 shows the frequency dependence of the median and P90 errors obtained with the 1D and 2D fits. Within the sub-LH window, both fits give errors below the $10^{-2}$ level and have comparable accuracy. In the transition region, $0.9 \lesssim f / f_{\text{LH}} \lesssim 3$, the 1D-fit error rises to the $10^{-1}$ level, whereas the 2D fit error remains close to $10^{-2}$. This indicates that much of the residual error in this transition region can be removed by including explicit frequency dependence in the fit. At higher frequencies, both fits exhibit an abrupt jump in error from the $10^{-2}$ level to the $10^{1}$ level, and the error remains at this order thereafter. The 2D fit delays the onset of this jump but does not eliminate it. All four devices exhibit this behaviour, although the jump frequency depends on the device parameters. These results show that the error outside the sub-LH window does not grow smoothly with frequency. Instead, it undergoes a sudden transition at specific frequencies. This indicates that the loss of accuracy outside the sub-LH regime stems from missing physical structure in the reduced model, rather than from the limited flexibility of the empirical correction alone.

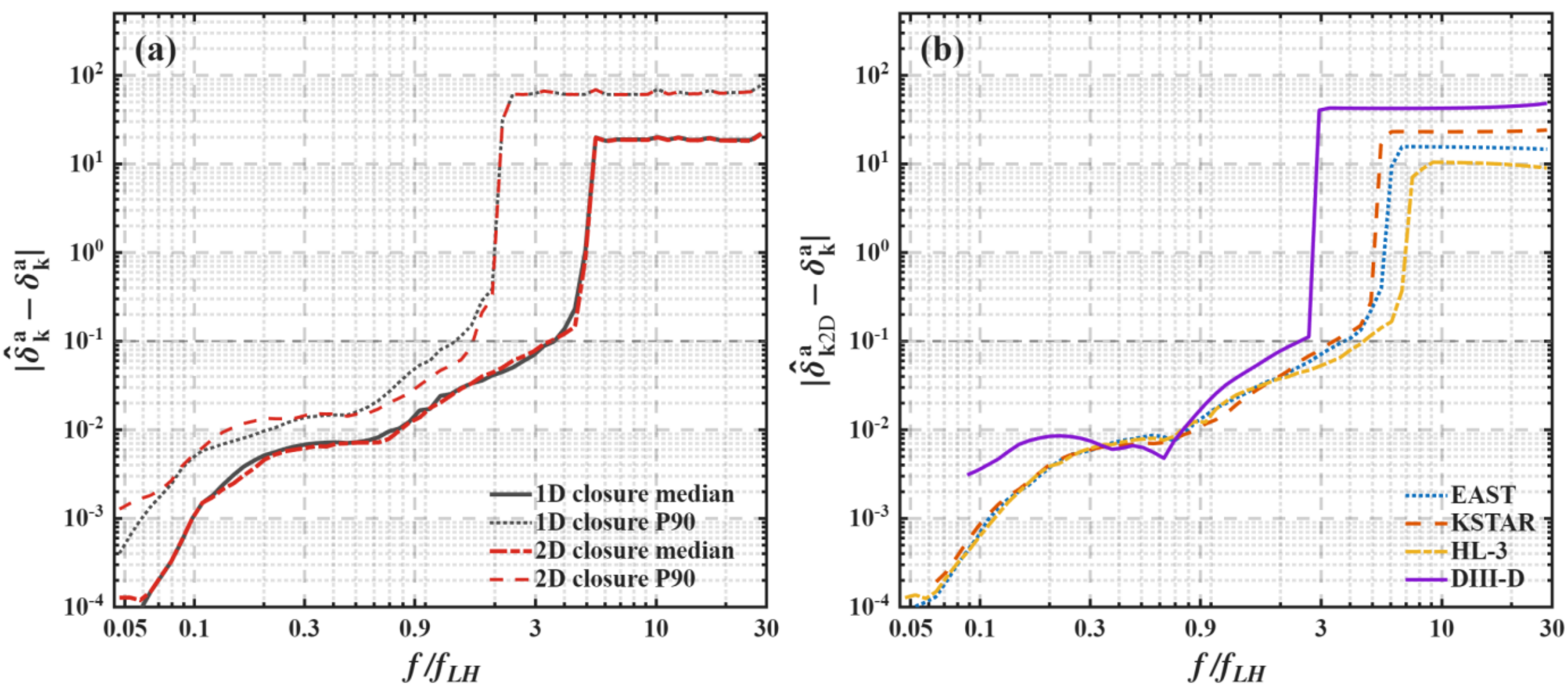

Figure 12. Frequency dependence of the residual errors obtained using the one-dimensional and two-dimensional empirical corrections in (a) the full dataset and (b) each device.

To identify the physical origin of the residual error jump, two characteristic frequencies are defined for each pair of $T_e$ and $N_{\parallel}$. The first $f_{\text{err}}$ denotes the frequency at which the prediction error first persistently exceeds the threshold of $10^{-1}$. The second $f_{\text{top}}$ denotes the frequency of the nearest abrupt change in the root structure of the cold-plasma dispersion relation around $f_{\text{err}}$. Figure 13 shows the correspondence between these two frequencies across the four devices. As shown in Figure 13, the data points are distributed close to the diagonal line over the entire frequency range of approximately $0.5-30 f_{\text{LH}}$, with no systematic deviation among the four devices. The median value of $|\log_{10}(f_{\text{top}}/f_{\text{err}})|$ is 0.031, and the P90 error is 0.170. These correspond to a median frequency separation of about 3.1% between the error jump and the nearest cold-dispersion topology event, with 90% of cases lying within a factor of approximately 17%. The close correspondence between $f_{\text{err}}$ and $f_{\text{top}}$ indicates that the abrupt increase in prediction error is strongly correlated with changes in the root structure of the cold-dispersion relation.

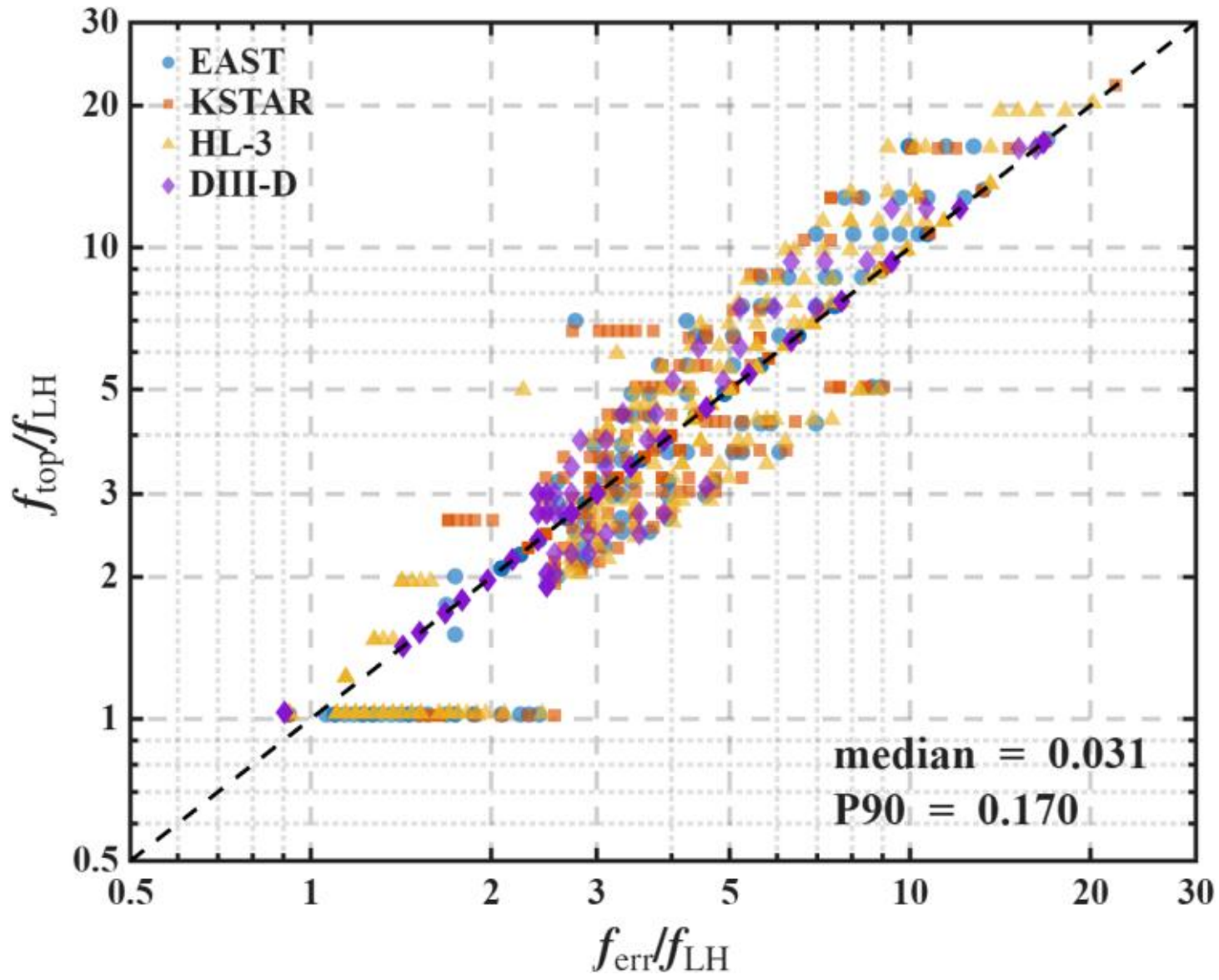

Figure 13. Correlation between the error-jump frequency and the nearest cold plasma dispersion topology-change frequency.

These results indicate that the residual error of the reduced model outside the sub-LH window is closely tied to the increasing complexity of the cold-plasma dispersion relation. In this regime, the physical structures neglected in the reduced model become increasingly important and progressively shrink the model's validity domain. The validity of the reduced model has been characterised from two complementary perspectives: its prediction accuracy and the boundaries of its applicability.

As a final check on the reference formulation itself, a branch-locked all-hot tensor comparison was performed in Appendix A. This test helps distinguish errors introduced by the reduced model itself from discrepancies already present between the Chiu-Chan reference and the higher-fidelity all-hot kinetic tensor. In the sub-LH range, the selected all-hot roots remain on the low-$N_\perp$ helicon/fast branch, and the median discrepancy between the Chiu-Chan model and the all-hot tensor is 4.9%. In the near-LH range, $1 < f / f_{LH} \le 5$, the median Chiu–Chan–all-hot discrepancy remains comparable at 5.3%. A clear degradation appears in the higher-frequency branch-clean range, $5 < f / f_{LH} \le 15$, where the median Chiu-Chan-All-hot discrepancy increases to 11.7% and the P90 error reaches 30.7%. This indicates a frequency-dependent kinetic-consistency boundary of the Chiu-Chan reference on the low-$N_\perp$ helicon/fast branch, rather than an artifact of root selection. At still higher frequencies, above approximately $16 f_{LH}$, the all-hot root identity becomes increasingly mixed. This range is therefore treated as a root-identity diagnostic rather than a clean helicon-branch benchmark. The next section examines how these errors, together with the empirical correction proposed here, propagate into predictions of current-drive efficiency.

# 5 Implications and Validation of Current Drive Model

### 5.1 Transfer of damping error to absorbed-power and CD source estimates

Since engineering applications of helicon waves typically involve both electron heating and non-inductive current drive, validating only the damping-rate accuracy does not suffice to establish the practical value of the reduced model. For current-drive applications, an error in the damping rate alters the local absorbed power and can therefore affect the estimated current-drive performance. It is thus necessary to further examine how the damping error of the reduced model propagates into current-drive-related quantities.

Ray-tracing codes such as GENRAY commonly employ the Ehst-Karney model to estimate the current-drive efficiency of helicon waves[32]. Following the structure of this type of model, and focusing on the error component directly tied to the damping prediction, we define the single-pass optical depth for the reference model as:

$$\tau^{\mathrm{ref}} = 2k_{\perp i}^{\mathrm{ref}} L \tag{69}$$

Since full ray-path integration is not performed in the present work, the effective single-pass propagation length is taken as L=a, where a is the minor radius of each device. This choice provides a device-scale estimate of the single-pass absorption thickness. Appendix B examines the sensitivity to this assumption by considering L=0.5a, a, and 2a, showing that the main conclusion is not dependent on L. The corresponding single-pass absorption fraction is:

$$f_{\mathrm{abs}}^{\mathrm{ref}} = 1 - \exp(-\tau^{\mathrm{ref}}) \tag{70}$$

Under the approximation that only the propagation of the damping error is considered, the current-drive term can be written as:

$$I_{\mathrm{CD}} \propto G(\xi_e, Z_{\mathrm{eff}})\left[1 - \exp(-\tau)\right] \tag{71}$$

where $G(\xi_e, Z_{\mathrm{eff}})$ denotes the weighting factor associated with the current-drive efficiency. The model error of $G$ itself is not considered here, only the error in the absorption term caused by the error in $k_{\perp i}$ is evaluated. A first-order expansion of $1 - \exp(-\tau)$ gives:

$$\delta_{\mathrm{CD}} = \left| \frac{\Delta I_{\mathrm{CD}}}{I_{\mathrm{CD}}^{\mathrm{ref}}} \right| \approx \mathcal{A}(\tau^{\mathrm{ref}}) \,|\, \delta_k \,| \tag{72}$$

with

$$\mathcal{A}(\tau^{\mathrm{ref}}) = \frac{\tau^{\mathrm{ref}} \exp(-\tau^{\mathrm{ref}})}{1 - \exp(-\tau^{\mathrm{ref}})} \tag{73}$$

represents the current-drive prediction error propagated from the damping-rate error.

The dependence of $\mathcal{A}$ on $\tau^{\mathrm{ref}}$ is shown in Figure 14(a). The curve separates two limiting regimes. On the weak-absorption side, where the reference single-pass optical depth is far below unity, $\mathcal{A}$ remains close to one, so the damping-rate error is transferred almost directly to the current-drive source estimate. On the strong-absorption side, where the reference optical depth is much larger than unity, the single-pass absorbed fraction approaches saturation and $\mathcal{A}$ decreases toward zero. In this regime, a finite damping-rate error produces only a weak additional change in the current-drive source estimate. Figures 14(b) and (c) show the distributions of $\delta_{\mathrm{CD}}$ for the uncorrected and corrected reduced models, respectively. The guide lines correspond to representative damping-error levels of $|\delta_k| = 0.05$, 0.10, and 0.30. For the uncorrected reduced model, most values of $\delta_{\mathrm{CD}}$ lie at the $10^{-1}$ level, showing that the damping error transfers substantially to the application-relevant source term. By contrast, after correction, the scatter points shift week downward and concentrate in a lower-error region. This indicates that the empirical correction not only reduces the damping prediction error at the $k_{\perp i}$ level but also weakens its transfer to the current-drive source term.

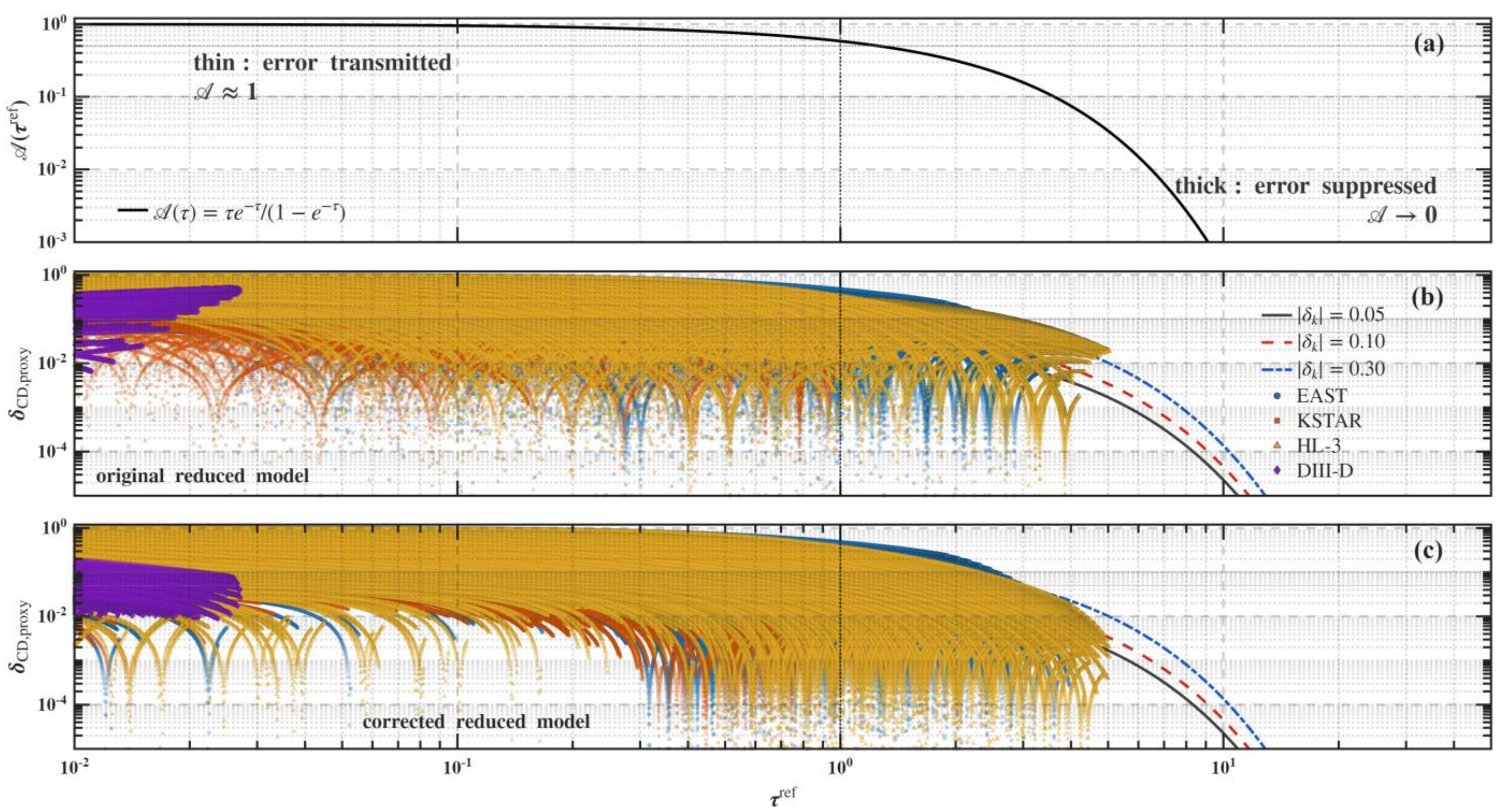

Figure 14. Current drive error induced by the damping error of the reduced model: (a) shows the transfer factor $\mathcal{A}(\tau^{\mathrm{ref}})$; (b) and (c) show the propagated source-proxy error $\delta_{\mathrm{CD}}$ for the original reduced model and the corrected reduced model.

These results confirm that the empirical correction developed above is meaningful at the level of application-relevant outputs. In the weak-absorption regime, the accuracy of $k_{\perp i}$ must be tightly controlled, because the damping error transfers almost directly. In the strong-absorption regimes, absorption saturation partly suppresses this transfer and reduces the sensitivity of the current-drive prediction to errors in $k_{\perp i}$.

### 5.2 Local-efficiency error within the heating-validity domain

To assess the applicability of the reduced current-drive source term, the error of $\tilde{\eta}_{\mathrm{red}}^{\mathrm{LD}}$ is evaluated within the domain. The domain is defined by the validity of the corrected heating model, namely $|\delta_k^{\mathrm{corr}}| \le 0.10$, coverage greater than 90%. Figure 15(a) shows the Case A results within this Domain. A total of 667 in-domain samples are obtained from the four devices. The median absolute error is 10.8%. Using a Landau-resonant absorption proxy weight, $W_i \propto \exp(-\xi_e^2)$, to represent the power-deposition weighting, the source-weighted mean absolute error defined by Eq.(74) is 10.1%, and the source-weighted P90 error is 20.1%.

$$E_\eta^{(w)} = \frac{\sum_i W_i \left|\eta_{\mathrm{red},i} - \eta_{\mathrm{ref},i}\right|}{\sum_i W_i \left|\eta_{\mathrm{ref},i}\right|} \times 100\% \tag{74}$$

Both the local median error and the source-weighted mean error remain at the $\sim 10\%$ level. The source-weighted P90

error remains below 20.1%, indicating that, for a prescribed absorbed-power profile, the reduced model reproduces the Ehst-Karney current-drive efficiency with an expected error at the order of 10%. Therefore, $\tilde{\eta}_{\mathrm{red}}^{\mathrm{LD}}$ can be used as an LD-consistent current-drive source proxy for rapid estimates of the mapping from $P_{\mathrm{abs}}$ to $J_{\mathrm{CD}}$.

Figure 15(b) shows the sensitivity of the reduced current-drive efficiency to the choice of reference model at the nearest-grid nominal point of each device. Circles and squares denote the signed deviations relative to the Landau-channel reference and the Alfvén-wave/TTMP reference from the Ehst-Karney parameterization, respectively. For all four devices, the deviation relative to the Alfvén-wave/TTMP reference is larger in magnitude than that relative to the Landau-channel reference, showing that the inclusion of the Alfvén-wave/TTMP contribution changes the estimated local current-drive efficiency. This nominal-point comparison is used as a local diagnostic of reference-model sensitivity, whereas the physically consistent validation of the reduced source term remains the like-for-like comparison with the Landau-channel reference.

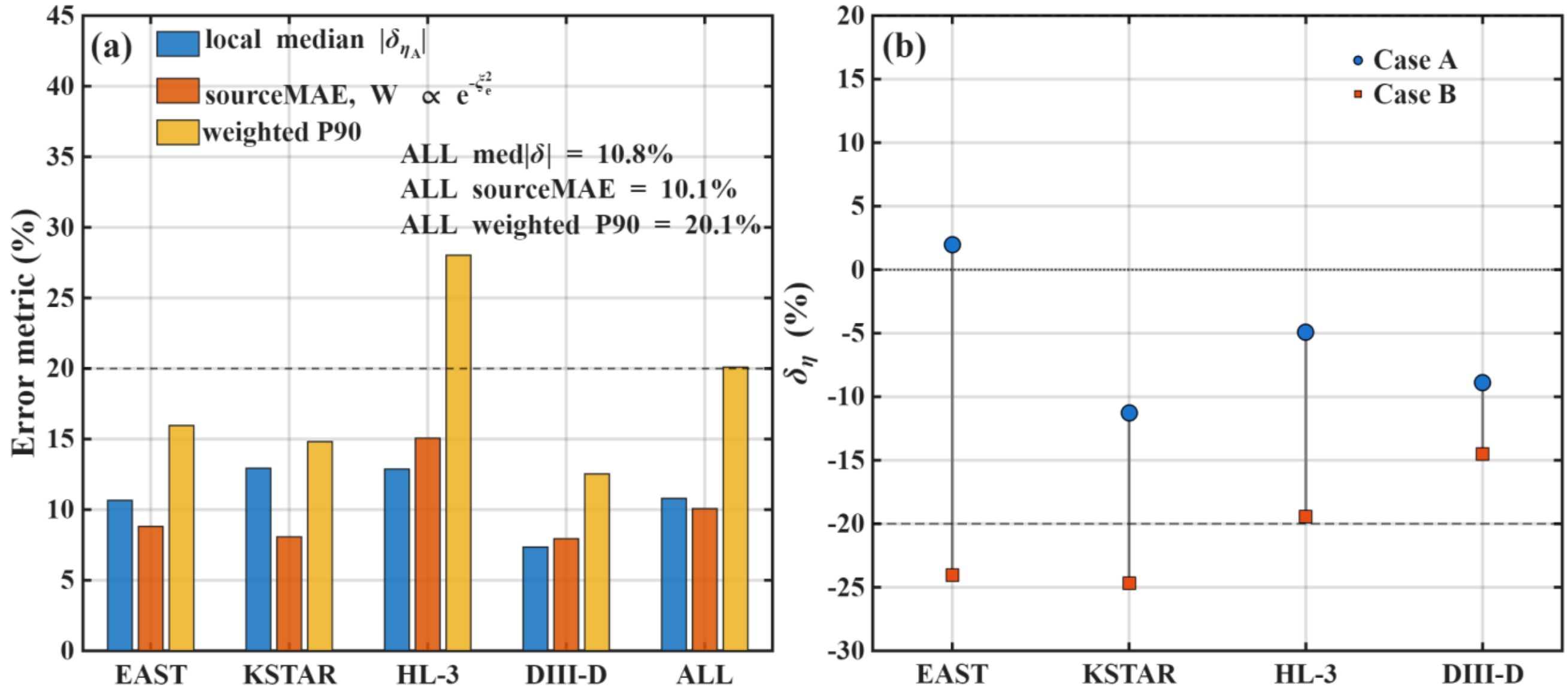

Figure 15. Reduced current-drive efficiency error within the heating-validity domain: (a) Case-A error statistics relative to the Ehst-Karney LD reference; (b) Nominal-point deviations relative to the Ehst-Karney LD and AW/TTMP references.

Table 3 further lists the device-resolved statistics. For Case A, DIII-D gives the lowest local error, with a median absolute error of 7.4%, whereas HL-3 and KSTAR give the largest error, 12.9%. The corresponding value for EAST is 10.7%. Thus, within the LD-consistent comparison range, the reduced model provides local current-drive-efficiency estimates with errors of approximately 7-15%. The absorption-weighted source-level errors remain at the 10% level.

The device-to-device variation in the Case-A error mainly reflects the different geometric and phase-velocity dependences of the trapped-particle correction used in the present reduced model and the R factor in the Ehst-Karney LD reference. For EAST, HL-3, and DIII-D, the deviation of the reduced model from the Ehst-Karney LD reference remains relatively small. HL-3 shows a somewhat larger Case-A deviation, indicating that the LD-consistent proxy is more sensitive to the trapped-particle closure for this superconducting, relatively high-field parameter set. This residual bias should therefore be interpreted as a limitation of the simplified trapping correction, rather than as an error caused by the omission of non-Landau absorption channels.

Table 3 also shows that the Case-B errors are systematically larger than the Case-A values. The all-device local median absolute error increases to 23.1%, and the absorption-weighted source-level MAE increases to 24.0%. This increase is expected because the Case-B reference includes the AW/TTMP branch of the Ehst-Karney parameterization, whereas the present reduced model retains only the LD channel. The Case-B deviation therefore reflects the change in the current-drive reference after the AW/TTMP contribution is included, not an error of the LD-only source term in a like-for-like comparison.

The P90 value is more strongly affected by a small subset of outlying points. EAST, DIII-D, and KSTAR show moderate Case-B P90 errors of 27.4%, 27.2%, and 36.8%, respectively, whereas HL-3 gives a much larger value of 178.7%. These large HL-3 deviations occur for samples close to the AW/TTMP-reference cutoff and increase the all-device P90 error to 96.8%. The median error remains 23.1%, so the central error level of Case B is still around 20%–25%. On this basis, the reduced current-drive source should be applied as an LD-channel model, and its physically consistent validation remains the Case-A comparison.

Table 3. Device-resolved errors of the reduced current-drive efficiency within the heating-validity domain.

| Device | Case A | | | Case B | | |
|---|---|---|---|---|---|---|
| | med-abs | src-MAE | P90 | med-abs | src-MAE | P90 |
| EAST | 0.107 | 0.088 | 0.160 | 0.244 | 0.167 | 0.274 |
| HL-3 | 0.129 | 0.151 | 0.280 | 0.253 | 0.548 | 1.787 |
| DIII-D | 0.074 | 0.079 | 0.125 | 0.146 | 0.121 | 0.272 |
| KSTAR | 0.129 | 0.081 | 0.148 | 0.269 | 0.192 | 0.368 |
| ALL | 0.108 | 0.101 | 0.201 | 0.231 | 0.240 | 0.968 |

**5.3 CFETR profile-level reconstruction under a prescribed absorption/ray input**

To further assess the applicability of the reduced model to current-drive calculations, this section uses the published CFETR helicon current-drive cases of Wu et al.[23] as reference cases. In that work, the current-density profiles were obtained from GENRAY/CQL3D calculations, where CQL3D is a Fokker-Planck solver used to calculate the electron distribution function and the resulting driven current[46]. Here, the published absorbed-power profile, $P_{\mathrm{abs}}^{\mathrm{pub}}(\rho)$, is prescribed. The purpose of this test is to examine whether the local reduced current-drive weighting, $\eta_{\mathrm{CD}}^{\mathrm{red}}(\rho)$, can reasonably reconstruct the radial shape of the CQL3D current-density profile, $J_{\mathrm{CD}}^{\mathrm{CQL3D}}(\rho)$.

To separate the dominant role of the absorbed-power profile from the modulation introduced by the local current-drive weighting, two reconstruction models are used. The first model uses only the published absorbed-power profile,

$$\hat{J}_{\mathrm{CD}}^{\mathrm{abs}}(\rho) = A_{\mathrm{abs}} P_{\mathrm{abs}}^{\mathrm{pub}}(\rho) \tag{75}$$

This absorption-only reconstruction tests how much of the radial structure of $J_{\mathrm{CD}}^{\mathrm{CQL3D}}$ is already determined by the absorbed-power profile. The second model further includes the local reduced current-drive weighting,

$$\hat{J}_{\mathrm{CD}}^{\eta}(\rho) = A_{\eta} \eta_{\mathrm{CD}}^{\mathrm{red}}(\rho) P_{\mathrm{abs}}^{\mathrm{pub}}(\rho) \tag{76}$$

Here, $\eta_{\mathrm{CD}}^{\mathrm{red}}$ is calculated using the reduced current-drive source model described in Section 2.2. The constants $A_{\mathrm{abs}}$ and $A_{\eta}$ are global amplitude-normalization factors obtained from least-squares fitting. They are used only to remove amplitude uncertainties associated with digitization, radial normalization, flux-surface volume factors, and absolute power normalization.

The post-processing comparison is shown in Figure 16 for two published cases from Wu et al.[23]. Figure 16(a) corresponds to the case shown in their Figure 4, and Figure 16(b) corresponds to the case shown in their Figure 5. The black dashed curves denote the digitized CQL3D current-density profiles. The yellow curves denote the absorption-only reconstruction, and the purple curves denote the efficiency-weighted reconstruction. For the Figure 4 case of Wu et al. [23], the absorption-only reconstruction already reproduces the peak location and the overall radial shape of the CQL3D current-density profile well, with $R_{\mathrm{shape}}^2 = 0.994$ and $\Delta\rho_{\mathrm{peak}} = 0.005$. After the local reduced current-drive weighting is included, the efficiency-weighted reconstruction still shows good agreement, although the shape metrics are slightly lower, with $R_{\mathrm{shape}}^2 = 0.969$ and $\Delta\rho_{\mathrm{peak}} = 0.009$. This indicates that, in this case, the main radial structure of $J_{\mathrm{CD}}$ is primarily determined by the absorbed-power profile, while the local current-drive weighting provides only a secondary modulation. For the Figure 5 case of Wu et al.[23], the absorption-only reconstruction also reproduces the CQL3D profile well, with $R_{\mathrm{shape}}^2 = 0.985$ and $\Delta\rho_{\mathrm{peak}} = 0.008$. After the reduced current-drive weighting is included, the efficiency-weighted reconstruction improves the agreement further, with $R_{\mathrm{shape}}^2 = 0.993$ and $\Delta\rho_{\mathrm{peak}} = 0.006$. This shows that, in this case, the radial variation of the local current-drive efficiency provides a useful correction to the reconstructed current-density profile.

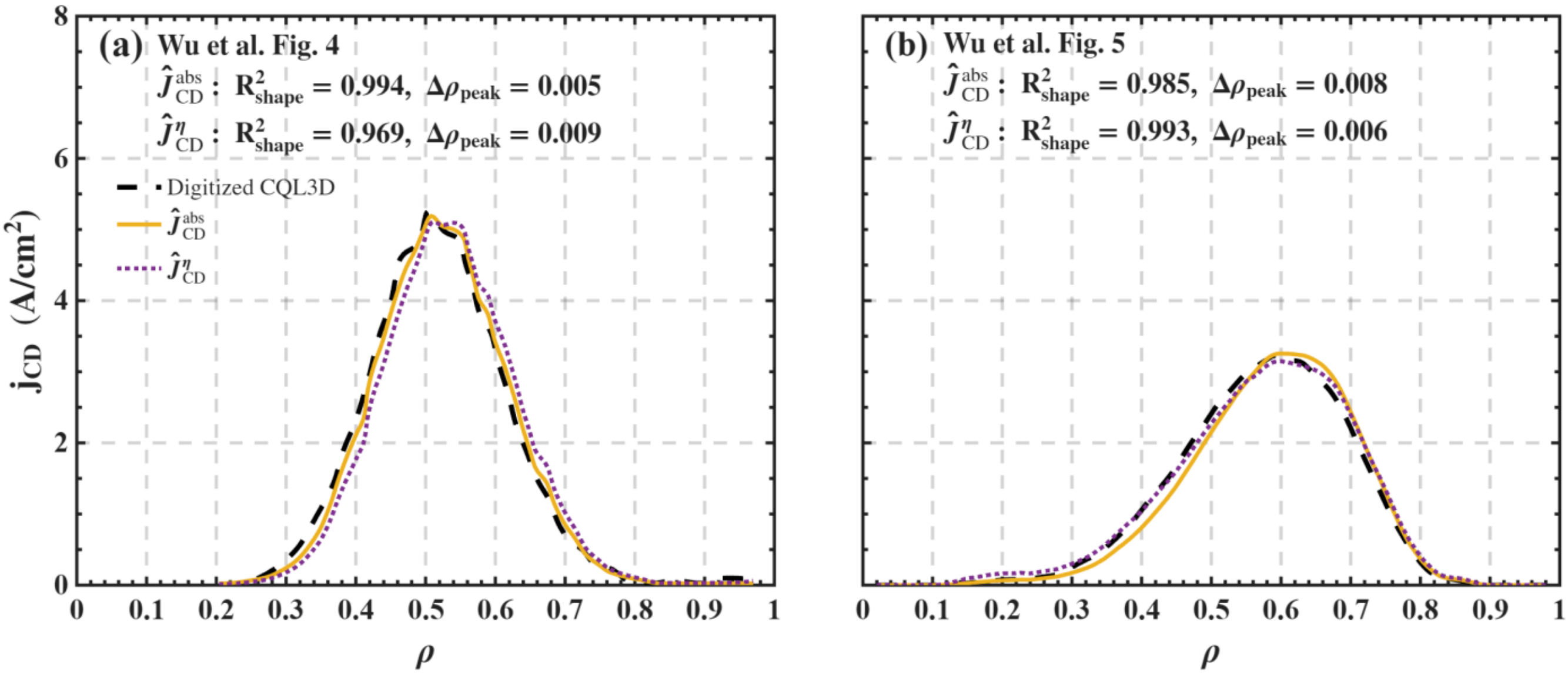

Figure 16. Post-processing reconstruction of published CFETR helicon current-density profiles using the reduced current-drive source.

Overall, both published cases show that, when $P_{abs}^{pub}(\rho)$ is prescribed, the main radial structure of the CQL3D current-density profile can be reasonably reconstructed from the absorbed-power profile, with a secondary modulation from the local current-drive weighting. The strong agreement obtained with the absorption-only reconstruction confirms that the absorbed-power deposition location is the dominant factor controlling the $J_{CD}$ profile. The efficiency-weighted reconstruction retains good shape-level agreement in both cases and further improves the profile reconstruction in the Figure 5 case. It should be emphasized that this section does not claim that the reduced model can independently reproduce the GENRAY/CQL3D results without ray-tracing information. This test only demonstrates that, within a post-processing framework with a prescribed published absorbed-power profile, the reduced current-drive source term provides a reasonable radial weighting.

# 6 Conclusion

This work developed and benchmarked a reduced model for tokamak helicon-wave heating and current drive in the ELD-dominated operating regime. The model retains only the parallel electron Landau damping channel and combines this kinetic damping term with the cold-plasma dispersion relation. The Maxwellian Landau absorptive correction is introduced only through the parallel response, so that the damping rate and the local current-drive efficiency can be evaluated without solving a thermally corrected hot-dispersion relation. This construction preserves the dominant helicon absorption physics in the intended regime while keeping the model simple enough for large parameter scans and ray-tracing-oriented workflows.

The heating model was benchmarked against the Chiu-Chan reference model over parameter scans based on EAST, HL-3, DIII-D and KSTAR. The scan shows that the reduced model has a well-defined validity domain in the sub-LH helicon window. The error decomposition separates the loss of non-Landau damping channels from the residual error within the retained Landau channel. In the Landau-dominated region, the residual factor from different devices collapses primarily onto a single function of $Z_R$. This behaviour indicates that the main remaining error is associated with the Hermitian thermal-response feedback omitted from the reduced dispersion relation. A fifth-order correction factor $\rho(Z_R)$ was therefore introduced to compensate for this missing response without requiring an additional hot-root solve. The corrected model substantially enlarges the low-error coverage and gives ITER-like and BEST-like residuals at the training-set level, supporting its use for extrapolative scans within the same helicon validity window. The extended frequency analysis identifies the upper boundary of this reduced description. Above the sub-LH range, both the uncorrected reduced model and the one-dimensional $\rho(Z_R)$ correction show abrupt error growth. This behaviour correlates with changes in the cold-dispersion branch structure and with the appearance of a high- $N_\perp$ branch near the low- $N_\perp$ helicon branch. Additional branch-locked all-hot diagnostics show that the initial super-LH error growth is not simply caused by a clean jump to the high- $N_\perp$ branch. At higher frequencies, however, branch proximity and branch-identity ambiguity become significant, especially for EAST and HL-3. These results indicate that the reduced model should be applied within the sub-LH helicon window, while the super-LH region requires an explicit treatment of branch selection, branch coupling and possible mode conversion.

For current drive, the corrected heating model was coupled to a reduced local current-drive source term and benchmarked against the Ehst-Karney parameterization. Within the heating-model validity domain, the like-for-like Landau-channel comparison gives a device-aggregated median absolute error of 10.8%, an absorption-weighted source-level MAE of 10.1%, and a P90 error of 20.1%. The device-resolved median errors range from 7.4% to 12.9%, showing moderate accuracy across the four parameter sets. A post-processing test using published absorbed-power profiles further shows that the reduced source term can reasonably reproduce the radial shape of CFETR GENRAY/CQL3D helicon current-density profiles. The prescribed absorption profile provides the dominant radial structure, while the local current-drive efficiency acts as a secondary weighting. The error-transfer analysis further shows that damping-rate errors do not propagate uniformly into absorbed power and current drive. In weak single-pass absorption, the damping error is transferred almost linearly to the current-drive source. In stronger absorption, saturation suppresses this transfer. The empirical heating correction therefore reduces the dominant source-level error to a few-percent scale in the tested propagation cases.

The present model is intended as a reduced closure for fast computations of helicon heating and Landau-channel current-drive estimates, not as a replacement for full-wave or hot-dispersion modelling outside its validity range. Future work will apply the corrected reduced model as a fast screening tool for DEMO-class reactor concepts. Within ray-based or integrated modelling workflows, the model can support large parameter scans over wave frequency, launched parallel refractive index, plasma profiles and magnetic-field conditions at low computational cost. These scans will be used to identify feasible helicon-wave heating and current-drive scenarios for different DEMO-relevant reactor designs, including the accessible operating window, deposition location, single-pass absorption and off-axis current-drive capability. The reduced model can therefore serve as a pre-selection tool before more expensive full-wave, ray-tracing/Fokker-Planck or experimental validation studies are performed.

## Acknowledgements

This work is supported by the National Natural Science Foundation of China (Nos. 92271113, 12411540222, 12481540165), the Natural Science Foundation Project of Chongqing (No. CSTB2025NSCQ-GPX0725), and the ENN's Hydrogen-Boron Fusion Research Fund (No. 2025ENNHB01-011).

## Appendix A: Auxiliary all-hot tensor benchmark

In the main text, the Chiu-Chan model is used as the primary reference model to evaluate the error structure and empirical correction of the reduced model. To examine whether the improvement achieved by the corrected model merely arises from the algebraic self-consistency within the damping factor of the Chiu-Chan model itself, this appendix further employs a complex-root calculation based on a fully hot dielectric tensor as an auxiliary validation of the reduced model.

### A.1 All-hot dielectric tensor and numerical implementation

In the local magnetic-field coordinate system, the background magnetic field is taken to be along the $z$ direction, and the wave vector is written as $\mathbf{k}=(k_\perp,0,k_\parallel)$. In this coordinate system, the hot dielectric tensor takes the form:

$$\varepsilon_{hot}=\begin{pmatrix}\varepsilon_{11} & \varepsilon_{12} & 0\\ -\varepsilon_{12} & \varepsilon_{22} & \varepsilon_{23}\\ 0 & -\varepsilon_{23} & \varepsilon_{33}\end{pmatrix} \tag{A1}$$

For prescribed $N_\parallel$, $\omega$, and local plasma parameters, the hot dispersion relation is solved directly for the complex perpendicular refractive index:

$$N_\perp = N_{\perp r}+iN_{\perp i} \tag{A2}$$

The corresponding all-hot damping rate is defined from the imaginary part of this root as:

$$k_{\perp i}^{\text{hot}}=\frac{\omega}{c}\left|\operatorname{Im} N_\perp\right| \tag{A3}$$

The hot dielectric tensor used here retains the ion cyclotron harmonic contributions, ion finite-Larmor-radius effects, the parallel electron Landau response, and the transverse correction terms associated with electron damping. The tensor elements used in the calculation are:

$$\varepsilon_{11}=1+\frac{\omega_{pe}^2}{\Omega_e^2}+\frac{\omega_{pi}^2}{\omega^2}\xi_{i0}\sum_{\ell=-L}^{L}\frac{\ell^2}{\mu_i}\Gamma_\ell(\mu_i)Z(\xi_{i\ell}) \tag{A4}$$

$$\varepsilon_{12}=-i\left[\frac{\omega_{pi}^2}{\omega^2}\xi_{i0}\sum_{\ell=-L}^{L}\ell\left[\Gamma_\ell(\mu_i)-\Gamma_{\ell'}(\mu_i)\right]Z(\xi_{i\ell})+\frac{\omega_{pe}^2}{\omega\Omega_e}\right] \tag{A5}$$

$$\varepsilon_{22}=\varepsilon_{12}+\frac{\omega_{pi}^2}{\omega^2}\xi_{i0}\sum_{\ell=-L}^{L}2\mu_i\left[\Gamma_\ell(\mu_i)-\Gamma_{\ell'}(\mu_i)\right]Z(\xi_{i\ell})+\chi_{e2} \tag{A6}$$

$$\varepsilon_{33}=1-\frac{\omega_{pi}^2}{\omega^2}+\frac{2\omega_{pe}^2}{\omega^2}\xi_e^2\left[1+\xi_e Z(\xi_e)\right] \tag{A7}$$

$$\varepsilon_{23}=\frac{i}{2}k_\perp\rho_e\frac{1}{\xi_e}\varepsilon_{33} \tag{A8}$$

with

$$\begin{aligned}&\chi_{e2}=\frac{\omega_{pe}^2}{\omega^2}k_\perp^2\rho_e^2\xi_e Z(\xi_e)\\ &\Gamma_\ell(\mu_i)=e^{-\mu_i}I_\ell(\mu_i)\qquad \mu_i=\frac{k_\perp^2\rho_i^2}{2}\\ &\xi_{i\ell}=\frac{\omega-\ell\Omega_i}{k_\parallel v_{ti}}\qquad \xi_e=\frac{\omega}{k_\parallel v_{te}}\end{aligned} \tag{A9}$$

Here, $I_\ell$ denotes the modified Bessel function of the first kind. For each local parameter point, the damping rate of the corrected reduced model is given by:

$$k_{\perp i}^{\text{corr}}=\frac{k_{\perp i}^{\text{red}}}{\rho_{\text{fit}}(Z_R)} \tag{A10}$$

where $\rho_{\text{fit}}(Z_R)$ is the empirical correction function obtained in the main text. The error relative to the complex-root result from the all-hot tensor is defined as:

$$\delta_{k,\text{hot}}^{\text{corr}}=\frac{\left|k_{\perp i}^{\text{corr}}-k_{\perp i}^{\text{hot}}\right|}{\left|k_{\perp i}^{\text{hot}}\right|} \tag{A11}$$

The low-error criterion is taken as $\left|\delta_{k,\mathrm{hot}}^{\mathrm{corr}}\right| \le 0.2$. For each two-dimensional parameter case, the low-error coverage is defined as

$$C_{\mathrm{hot}} = \frac{N\left(\left|\delta_{k,\mathrm{hot}}^{\mathrm{corr}}\right| \le 0.2\right)}{N_{\mathrm{valid}}} \times 100\% \quad \text{(A12)}$$

where $N_{\mathrm{valid}}$ denotes the number of cases in that case for which a single cold branch, a valid reduced-model damping rate, and a converged complex root of the all-hot tensor are all obtained.

**A.2 Sub-LH consistency with the all-hot complex-root result**

The first all-hot comparison is restricted to the same sub-LH single-branch window used in the main text. This test is intended to check whether the $Z_R$-based correction, which was calibrated against the Chiu-Chan reference model, also remains consistent with a more direct complex-root calculation based on the hot dielectric tensor. For each valid scan point, the corrected reduced damping rate is compared with $k_{\perp i}^{\mathrm{hot}}$, and the low-error coverage is evaluated using the criterion $\left|\delta_{k,\mathrm{hot}}^{\mathrm{corr}}\right| \le 0.2$.

Figure A1 shows the low-error coverage of the corrected reduced model relative to the all-hot tensor calculation. The low-error region obtained from the all-hot tensor is broadly consistent with the coverage structure obtained in the main text using the Chiu-Chan reference. This agreement indicates that the correction proposed in this work is not merely an empirical compensation for the error relative to the Chiu-Chan damping formula, but also remains consistent with a more direct complex-root calculation based on the hot dielectric tensor.

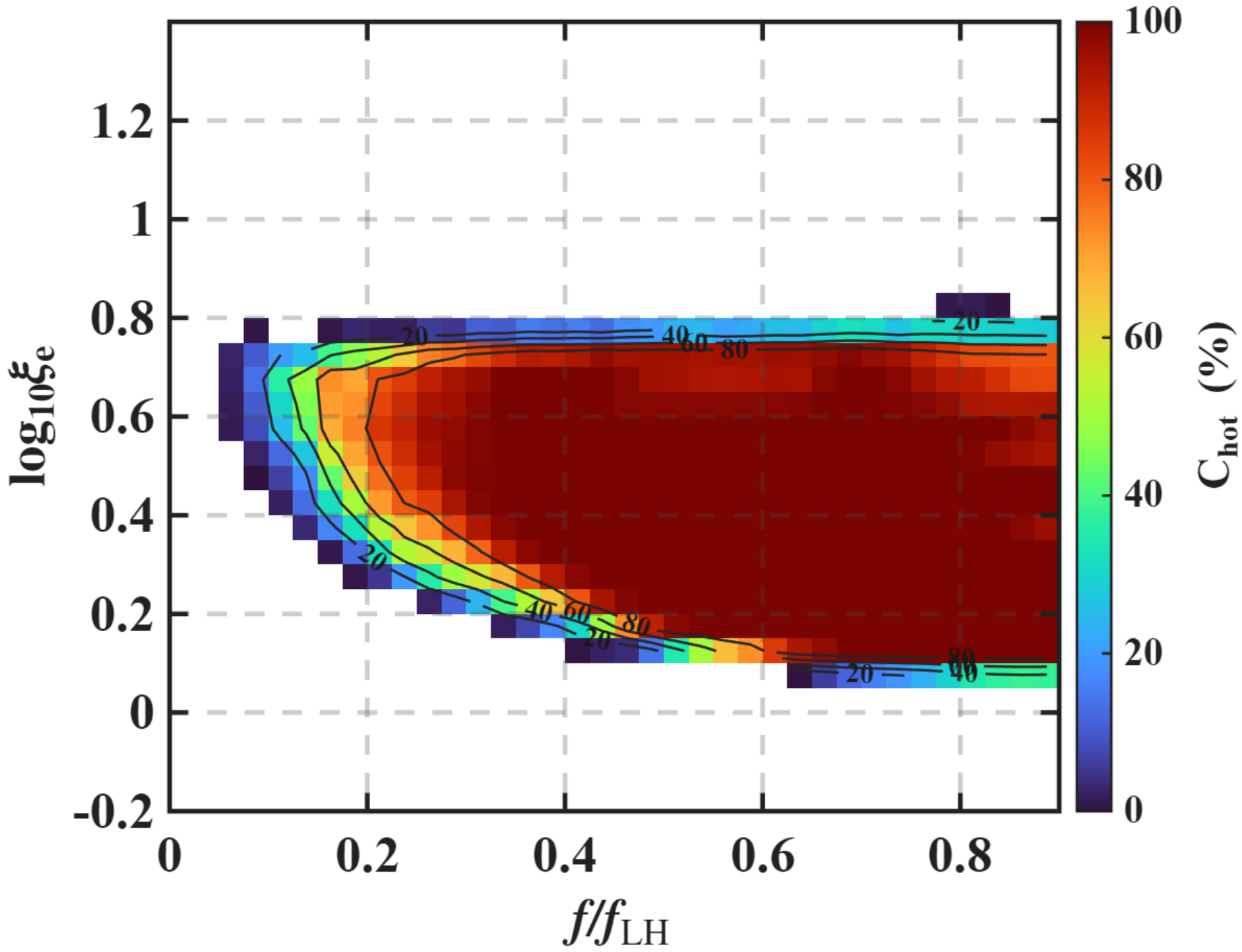

Figure A1. The low-error coverage of the corrected reduced model relative to the all-hot tensor calculation.

## Appendix B: Sensitivity of the current-drive error to the effective propagation length

In this study, the effective single-pass propagation length is taken as L=a to estimate the absorption thickness of the reference model and to analyze the effect of damping-rate errors on the current-drive error. It is therefore necessary to examine whether this assumption significantly affects the conclusions drawn in Section 5.

Table B1 summarizes the statistical results for the three choices of $L$. As shown in Table B1, increasing $L$ from 0.5a to 2a shifts the distribution of $\tau^{\mathrm{ref}}$ proportionally toward larger absorption thicknesses, with the P90 value increasing from 0.2 to 0.8. This indicates that a larger propagation length enhances the contribution from the high-absorption tail, although most cases remain in the weak- to moderately weak-absorption regime within the present parameter scan. Meanwhile, the median value of the error-transfer factor remains close to unity for all three choices of $L$. Its P10 value decreases from 0.903 to 0.652 as $L$ increases, indicating that absorption saturation begins to suppress error transfer in the high-absorption tail. However, this variation does not significantly change the statistics of the propagated current-drive proxy error of the corrected model. For all three choices of $L$, the median value of $\delta_{\mathrm{CD}}^{\mathrm{corr}}$ remains around 0.016, while the P90 error is approximately 0.196. Therefore, the main conclusion regarding the transfer of damping-rate errors to current-drive proxy errors does not depend on the specific choice of the effective single-pass propagation length. The effective propagation length changes the quantitative placement of cases in absorption-thickness space and affects the error-transfer factor in the high-absorption tail, but it does not alter the statistical assessment of the corrected model's current-drive proxy error.

Table B1. Sensitivity of the current-drive error-transfer estimate to the assumed effective propagation length

| Quantity | L=0.5a | L=a | L=2a |
| --- | --- | --- | --- |
| median $\tau^{ref}$ | $5.95\times10^{-4}$ | $1.19\times10^{-3}$ | $2.38\times10^{-3}$ |
| $P90(\tau^{ref})$ | 0.200 | 0.401 | 0.801 |
| median A | 0.9997 | 0.9994 | 0.9988 |
| $P10(A)$ | 0.903 | 0.813 | 0.652 |
| median $\delta_{CD}$ | 0.0171 | 0.0166 | 0.0160 |
| $P90(\delta_{CD})$ | 0.196 | 0.196 | 0.196 |

## Appendix C. Details of the Ehst-Karney parameterization and AW-reference diagnostics

For each damping channel, the Ehst-Karney model analytical efficiency is written as the product of a channel kernel and three correction factors:

$$\tilde{\eta}_{\mathrm{ref}} = CRM\eta_{\mathrm{ker}} \tag{C1}$$

Here, $C$ is the spectral factor, $R$ is the trapped-particle correction, $M$ is the channel-specific enhancement factor, and $\eta_{\mathrm{ker}}$ is the dimensionless kernel of the corresponding channel. Because the velocity normalization and the conversion between electron temperature and velocity are not written in the same form in the present work and in the Ehst-Karney model, the original expressions are first rewritten using a unified velocity normalization. For the LD channel of the Ehst-Karney model, the reference efficiency is

$$\tilde{\eta}_{\mathrm{ref}}^{\mathrm{LD}} = \left(\frac{3.0}{Z_{\mathrm{eff}}\, w_{\mathrm{EK}}} + \frac{3.83}{Z_{\mathrm{eff}}^{0.707}} + \frac{4w_{\mathrm{EK}}^2}{5+Z_{\mathrm{eff}}}\right)\cdot\left\{1-\exp\left[-\left(\frac{0.389(1-\epsilon)w_{\mathrm{EK}}^2}{\epsilon}\right)^{1.38}\right]\right\}\cdot\left(1-\frac{\epsilon^{0.77}\sqrt{3.5^2+w_{\mathrm{EK}}^2}}{3.5\epsilon^{0.77}+w_{\mathrm{EK}}}\right) \tag{C2}$$

where $\epsilon = a/R_0$, and $w_{\mathrm{EK}} = \sqrt{m_e c^2/T_e}\Big/N_{\parallel}$. For the AW channel of the Ehst-Karney model, the reference efficiency is:

$$\tilde{\eta}_{\mathrm{ref}}^{\mathrm{AW}} = \left(\frac{11.91}{(0.678+Z_{\mathrm{eff}})w_{\mathrm{EK}}} + \frac{4.13}{Z_{\mathrm{eff}}^{0.707}} + \frac{4w_{\mathrm{EK}}^2}{5+Z_{\mathrm{eff}}}\right)\cdot\left\{1-\exp\left[-\left(\frac{0.0987(1-\epsilon)w_{\mathrm{EK}}^2}{\epsilon}\right)^{2.48}\right]\right\}\cdot\left(1+\frac{12.3\epsilon^{3/2}}{w_{\mathrm{EK}}^3}\right)\cdot\left(1-\frac{\epsilon^{0.77}\sqrt{3.5^2+w_{\mathrm{EK}}^2}}{3.5\epsilon^{0.77}+w_{\mathrm{EK}}}\right) \tag{C3}$$

where

$$\eta_{\mathrm{ker}}^{\mathrm{AW}} = \frac{11.91}{(0.678+Z_{\mathrm{eff}})w_{\mathrm{EK}}} + \frac{4.13}{Z_{\mathrm{eff}}^{0.707}} + \frac{4w_{\mathrm{EK}}^2}{5+Z_{\mathrm{eff}}} \tag{C4}$$

$$C^{\mathrm{AW}} = 1-\exp\left[-\left(\frac{0.0987(1-\epsilon)w_{\mathrm{EK}}^2}{\epsilon}\right)^{2.48}\right] \tag{C5}$$

$$M^{\mathrm{AW}} = 1+\frac{12.3\epsilon^{3/2}}{w_{\mathrm{EK}}^3} \tag{C6}$$

$$R = 1-\frac{\epsilon^{0.77}\sqrt{3.5^2+w_{\mathrm{EK}}^2}}{3.5\epsilon^{0.77}+w_{\mathrm{EK}}} \tag{C7}$$